\documentclass[fleqn,usenatbib]{mnras}

\usepackage{newtxtext,amsmath}

\usepackage[T1]{fontenc}
\usepackage{ae,aecompl}
\usepackage[normalem]{ulem}

\usepackage{graphicx}	
\usepackage{amsmath}	
\usepackage{amssymb}	

\usepackage{xcolor}
\usepackage{threeparttable}

\def\ms{\mbox{$M_{\ast}$}}
\def\mha{\mbox{$M_{\rm H\,{\sc I}}$}}
\def\mhatot{\mbox{$M^{\rm tot}_{\rm H\,{\sc I}}$}}
\def\mhacen{\mbox{$M^{\rm cen}_{\rm H\,{\sc I}}$}}
\def\mhaSsat{\mbox{$M^{\rm sats}_{\rm H\,{\sc I}}$}}
\def\mhm{\mbox{$ M_{\rm H_{2}}$}}
\def\mh{\mbox{$M_{h}$}}
\def\mhp{\mbox{$M_{\rm DM}$}}
\def\vm{\mbox{$V_{\rm max}$}}
\def\vmax{\mbox{$V_{\rm DM}$}}
\def\vpeak{\mbox{$V_{\rm peak}$}}
\def\msun{\mbox{M$_{\odot}$}}

\def\HI{\mbox{H\,{\sc I}}}
\def\H2{\mbox{$\rm H_{2}$}}

\def\RHI{\mbox{$R_{\rm H\,{\sc I}}$}}
\def\RH2{\mbox{$R_{\rm H_{2}}$}}
\def\rv{\mbox{$R_{\rm vir}$}}%

\def\gsmf{\mbox{GSMF}}

\usepackage{url}
\newcommand{\rockstar}{\textsc{Rockstar}}

\newcommand{\smdp}{\texttt{SMDP}}
\usepackage{color} 

\def\mnras{\mbox{MNRAS}}
\def\apj{\mbox{ApJ}}
\def\aap{\mbox{A\&A}}
\def\apjl{\mbox{ApJ}}
\def\apjs{\mbox{ApJS}}
\def\aj{\mbox{AJ}}

\def\rmxaa{\mbox{RMxAA}}

\defcitealias{Yang+2012}{Y12}
\defcitealias{RAD13}{RAD13}
\defcitealias{RDA12}{RDA12}

\title[\HI--(sub)halo connection and \HI\ clustering]{The galaxy \HI--(sub)halo connection and the \HI\ spatial clustering of local galaxies}

\author[A. R. Calette et al.]{
A. R. Calette$^{1}$\thanks{E-mail: acalette@astro.unam.mx},
Aldo Rodr\'iguez-Puebla$^{1}$,
Vladimir Avila-Reese$^{1}$ \and 
and Claudia del P. Lagos$^{2,3}$ \\
$^{1}$Instituto de Astronom\'ia, Universidad Nacional Aut\'onoma de M\'exico, A. P. 70-264, 04510, Ciudad de M\'exico, M\'exico\\
$^{2}$ International Centre for Radio Astronomy Research (ICRAR), M468, University of Western Australia, 35 Stirling Hwy,\\ Crawley, WA 6009, Australia. \\
$^{3}$ ARC Centre of Excellence for All Sky Astrophysics in 3 Dimensions (ASTRO 3D). \\
}

\date{Accepted XXX. Received YYY; in original form ZZZ}

\pubyear{2021}

\begin{document}
\label{firstpage}
\pagerange{\pageref{firstpage}--\pageref{lastpage}}
\maketitle
%
\begin{abstract}
We extend the local stellar galaxy-(sub)halo connection to the atomic hydrogen (\HI) component by seeding semi-empirically galaxies into a large N-body dark matter (DM) simulation. The main input to construct the mock galaxy catalogue are: our constrained stellar mass-to-(sub)halo circular velocity (\ms--\vmax) relation, assuming a scatter independent of any galaxy property, and the empirical \mha{} conditional probability distributions given \ms\ for central and satellite galaxies. 
We find that the $\langle\log\mha\rangle-\log\mhp$ relation is not a monotonic increasing function. It increases with mass up to $\mhp\sim 10^{12}$ \msun, attaining a maximum of $\langle\log(\mha/\msun)\rangle \sim 9.2$, and at higher (sub)halo masses, $\langle\log(\mha)\rangle$ decreases slightly with \mhp. The scatter around it is also large and mass dependent. The bivariate  \mha{} and \mhp\ distribution is broad and bimodal, specially at $\mhp\gtrsim 10^{12}$ \msun, which is inherited from the input \mha\ conditional distributions. We also report the total (central+satellites) \HI\ gas mass  
within halos, \mhatot, as a function of \mhp. The mean $\mhatot$-$\mhp$ relation is an increasing monotonic function. The galaxy spatial clustering increases weakly as the \mha{} threshold increases. Our \HI\ mock galaxies cluster more in comparison to the blind \HI\ ALFALFA (Arecibo Fast Legacy ALFA) survey but we show that it is mainly due to the selection effects. We discuss the implications of our results in the light of predictions from semi-analytical models and hydrodynamics simulations of galaxy evolution. 
\end{abstract}




\section{Introduction}
\label{intro}

According to the current scenario of cosmic structure formation, galaxies formed and evolved within 
cold dark matter (CDM) halos from accreted and cooled gas within them. Depending on the different astrophysical processes and on the halo gravitational potential, a fraction of the cold accreted gas is deposited into the galaxy, part of this gas is transformed into stars while other part is heated or expelled by feedback processes  
due to stars and active galactic nuclei (AGNs). Therefore, the cold gas content that present-day galaxies have, given their halo and stellar masses (\mh\ and \ms, respectively), provides important information to constrain the different processes of galaxy evolution \citep[see e.g.,][and more references therein]{Crain+2017}. The cold (neutral) gas in the interstellar medium of galaxies is composed by atomic hydrogen (\HI), molecular hydrogen  (\H2), helium, and metals, being the dominant in mass the \HI\ component \citep[e.g.,][]{Fukugita+2004,Read+2005,Rodriguez-Puebla+2020}.

Much work has been done in studying the \HI\ content of galaxies in the nearby Universe. Among the largest efforts in this direction, we highlight the completion of the blind radio surveys \HI\ Parkes All-Sky Survey \citep[HIPASS;][] {Meyer+2004} and Arecibo Fast Legacy ALFA Survey \citep[ALFALFA;][]{Giovanelli+2005,Haynes+2011,Haynes+2018}. 
While blind radio HI surveys do not impose any selection criteria on the target sample, their shallow flux detection limits, \HI\ line width thresholds, and the small volumes surveyed bias the samples in other galaxy properties than the \HI\ mass. For example, it is clear that they are biased to \HI-rich, spiral blue galaxies \citep[e.g.,][]{Baldry+2008, Haynes+2011,Catinella+2012,Papastergis+2012,Maddox+2015,Chauhan+2019}. Moreover, these biases are mass dependent \citep[see e.g.,][]{Huang+2012,Maddox+2015}.
From these surveys, it was determined the local \HI\ mass function \citep[MF;][]{Zwaan+2005,Martin+2010,Papastergis+2012,Jones+2018}. While the total \HI\ MF seems robust against the biases mentioned above (down to the masses allowed by the 21-cm flux limit of the used radiotelescopes), this is not  the case for the \HI-to-stellar mass relation \citep[see e.g.,][]{Maddox+2015}, the projected two-point correlation function (2PCF) at different \HI\ mass thresholds \citep{Papastergis+2013,Guo+2017} and the \HI\ velocity function \citep{Chauhan+2019}. 
As a result of the biases, the \HI\ 2PCFs measured, for instance from the \HI\ ALFALFA survey, are expected to be lower than for all galaxies \citep{Martin+2012}. Indeed, the measurements show that for \HI-selected galaxies, the clustering amplitude is low and, depending on how the corrections to the biases are applied, different dependencies of clustering on \mha\ are found \citep[][]{Basilakos+2007,Meyer+2007,Martin+2012,Papastergis+2013,Guo+2017}. By means of a (sub)halo\footnote{Throughout this paper, we use the term (sub)halo to refer both for distinct haloes and subhaloes at the same time. A distinct or host halo is defined as a gravitational bounded sphere of radius \rv\ such that its overdensity at this radius is $\delta_{\rm vir}$ times the background density, 
and is not contained within a larger halo. A subhalo is a former halo whose centre is within the radius of a larger (distinct) halo and remains as a gravitational bounded system.} abundance matching model (SHAM), \citet[][]{Guo+2017} attempted to reproduce the clustering measurements of the ALFALFA sample by selecting only haloes that formed relatively late. This result is interpreted as an intrinsic assembly bias effect on the \HI\ gas content of galaxies. 
However, recent semi-analytical models of galaxy evolution do not find assembly age to be strongly correlated with the \HI\ content of galaxies \citep[][]{Spinelli+2020,Chauhan+2020}.

In the future, facilities such as the the Five-hundred-meter Aperture Spherical radio Telescope \citep[][]{Nan+2011,Li+2013}, the Square Kilometre Array \citep[SKA;][]{Blyth+2015,Carilli+2015}, or precursor instruments such as the Australian SKA Pathfinder \citep[][]{Johnston+2008} and the outfitted Westerbork Synthesis Radio Telescope, will bring extragalactic \HI\ studies more in line with optical/infrared surveys. The blind \HI\ surveys will then be much deeper and for larger volumes than the current ones, helping to reduce the strong sample selection effects. By now, the \HI-to-stellar mass and other correlations have to be constrained from radio follow-up observations of optically selected galaxy samples 
\citep[e.g.,][]{Catinella+2010,Catinella+2013,Catinella+2018,Wei+2010, Saintonge+2011,Papastergis+2012,Kannappan+2013,Boselli+2014,Eckert+2015,Stark+2016}. 

In \citet[][hereafter Paper I]{Calette+2018} we undertook the task of compiling and homogenizing from the literature as much as possible galaxy samples, in the spirit of the ones listed above (including most of them), with the additional requirement of information on the galaxy morphology because the HI gas content of galaxies strongly depends on morphology. For determining the \HI\ mass distributions, we have taken into account the reported upper limits for the radio non-detections:  after homogenizing them, in particular those from the distant galaxies of the \texttt{GASS} sample, we have applied a survival analysis. As a result, we were able to constrain not only the mean \mha--\ms\ and \mhm--\ms\ relations and their standard deviations for late- and  early-type galaxies (LTG and ETG, respectively) down to $\ms\sim 10^7$ \msun, but the respective full conditional probability density distribution functions (PDFs) of the \HI- and \H2-to-stellar mass ratios for a given \ms\ \citep[for previous attempts to constrain these distributions in the case of \HI\ see e.g.,][]{Lemonias+2013}. 

An important step in Paper I was {\it i)} to analyze separately late- and early-type galaxies given their very different gas distributions, {\it ii)} to take into account and treat adequately the \mha\ upper limits, which dominate in number for the massive ETGs, and
{\it iii)} to constrain the complete \mha\ and \mhm\ conditional PDFs given \ms. From these distributions, it is possible to calculate any statistics, for instance, the mean logarithmic or arithmetic \mha--\ms\ and \mhm--\ms\ relations.

Furthermore, in \citet[][hereafter Paper II]{Rodriguez-Puebla+2020} we derived and used a well-constrained Galaxy Stellar Mass Function (\gsmf) for all, late- and early-type galaxies down to $\sim 10^7$ \msun{}, and combined them with the \HI{} and \H2\ conditional PDFs to generate the bivariate or joint (\ms,\mha) and (\ms,\mhm) distribution functions. By projecting these bivariate distributions into the \HI\ and \H2\ axes, we obtained the \HI\ and \H2\ MFs, for LTGs and ETGs, as well as for all galaxies. In Paper II it was shown that our empirical \HI{} and \H2\ MFs agree well with those measured from blind radio surveys or optically selected radio samples corrected for volume incompleteness. In particular, we have shown that the \HI\ MF from blind radio surveys, like HIPASS and ALFALFA, are not affected by their selection effects. 
We concluded that our empirical statistical description of the local galaxy population regarding the stellar and \HI{} and \H2\ gas contents integrates and extends well a large body of observational information.

\subsection{The stellar--HI--dark matter mass connection}

The bivariate \ms\ and \mha\ distribution of the local galaxy population described above can be used to extend the empirical galaxy-halo connection to the \HI\ gaseous component of galaxies. 
There were some previous attempts to set the galaxy-halo connection for \mha, by means of the Halo Occupation Distribution (HOD) model and SHAM \citep[e.g.,][]{Papastergis+2012,Papastergis+2013,Guo+2017,Padmanabhan+2017,Paul+2018,Obuljen+2019}, or from physically motivated models that employed results from semi-empirical models \citep[see e.g.,][]{Popping+2015}.  
The results are quite diverse, both regarding the \HI-to-halo mass relation and its scatter, and the \HI\ galaxy clustering. A potential shortcoming in applying the SHAM for \mha\ is that the key assumption of this technique -the existence of a monotonic and tight relation between \mha{} and \mh{} (or any other halo property)- is probably not obeyed. On the other hand, the use of the measured \HI\ 2PCFs in the HOD models could lead to incorrect \HI-to-halo mass relations given the strong biases that the measured \HI\ 2PCFs suffer due to the sample selection effects mentioned above.
There are also some recent studies aimed at determining the \HI\ gas content of halos selected from optical galaxy group catalogues and by measuring the \HI\ stacked spectra of the entire groups \citep{Guo+2020} or by using the measured \HI\ masses of the galaxy members from blind \HI\ surveys \citep{Ai+2018,Tramonte+2020,Lu+2020}. 

The results from the semi-empirical studies mentioned above may be affected by several selection and confusion effects in the observational data that they use. On the other hand, they determine the total galaxy \HI\ mass within haloes but not the central (satellite) galaxy \HI\ mass in haloes (subhaloes).
Thus, there is still a lack of full and self-consistent empirical determinations of the galaxy stellar-\HI-(sub)halo mass connection, which is crucial for constraining theoretical models \citep[][]{Obreschkow+2016,Romeo2020,Romeo+2020} or for comparing with predictions from semi-analytical models and hydrodynamics simulations \citep[e.g.,][]{Kim+2017,Villaescusa-Navarro+2018,Diemer+2019,Baugh+2019,Spinelli+2020,Chauhan+2020}.

In order to establish the above-mentioned self-consistent connection, in this paper we take a different approach with respect to previous ones.  First, we perform a non-parametric SHAM with scatter in a large N-body cosmological simulation. The SHAM is applied to the stellar mass of galaxies and the maximum circular velocity of halos, \vm{}, and peak circular velocity of subhalos, \vpeak{}; the use of halo/subhalo velocities instead of masses allows us to reproduce the dependence of the 2PCF on {\it stellar mass}.
The SHAM result is used to assign stellar masses to the halos and subhalos in the simulation, and given \ms, the \HI\ mass is drawn from the empirical \HI-to-stellar mass distributions from Paper II. For the latter step, the \HI-to-stellar mass distributions of LTGs and ETGs are allowed to differ between centrals and satellites 
by using the functions constrained in \citet{Calette+2021} based on the recent extended GALEX Arecibo SDSS  Arecibo survey \citep[\texttt{xGASS;}][]{Catinella+2018}, where galaxies were separated into centrals and satellites.  In this way, we construct an empirically based mock (sub)halo-galaxy catalogue that allows us to predict the \HI-to-(sub)halo mass relation for central and satellite galaxies as well as the \HI\ spatial clustering.

In this paper, we are interested in making predictions of the
 stellar-\HI-halo mass connection for all the population of galaxies (LTGs+ETGs). 
In that regards, we take into account two relevant aspects: that the empirical \HI\ conditional distributions given \ms\ are different for central and satellite galaxies, and that to compare the results with the observational surveys, their biases must be considered. In particular, we explore the impact of these two aspects on the \HI\ spatial clustering and its comparison with the ALFALFA measurements. 
We find that in order to reproduce the (biased) measurements of the ALFALFA \HI\ clustering it will be necessary to implement more complex models to assign stellar and \HI\ masses to the (sub)haloes \citep[see e.g. e.g.,][]{Papastergis+2013,Guo+2017}. This exploration will be performed in a forthcoming paper. However, it is important to note that our prediction of the \mha\ distribution in (sub)haloes for {\it all} galaxies is insensitive to these modifications.

The outline of this paper is as follows. In Section \ref{methodology} we present our  methodology and the inputs needed to construct a mock catalogue with galaxies (halos) information. In Section \ref{results} we show our results from the mock catalogue generated such as the extended galaxy-(sub)halo connection, the \HI\ spatial clustering, and some consistency tests. We discuss these results in Section \ref{discussion} and compare some of them with theoretical predictions. Finally, we give our conclusions and future work in Section \ref{conclusions}.

\begin{figure*}
\includegraphics[height=5.4in,width=7in]{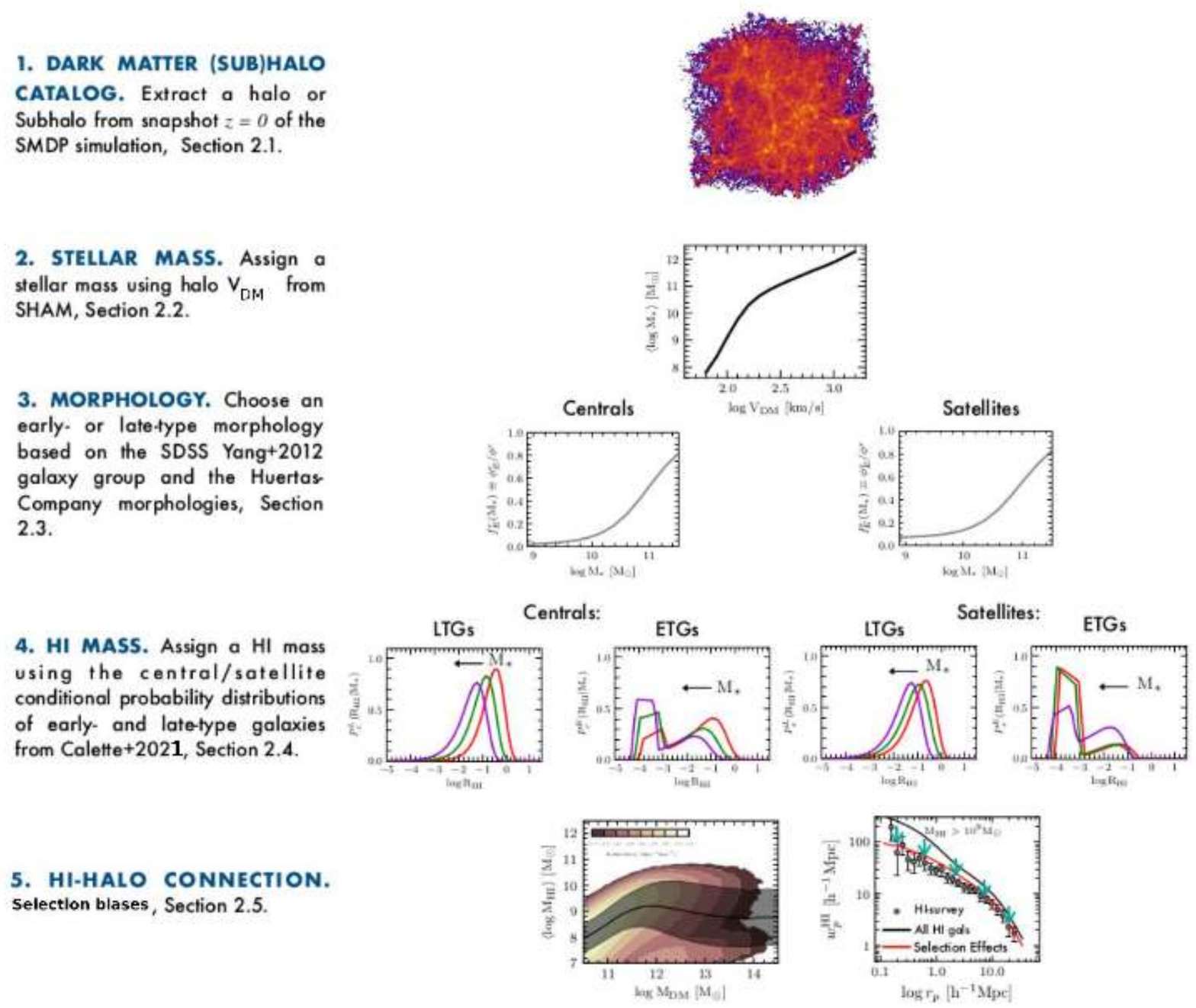}
\caption{Schematic figure for the five steps taken to implement our mock galaxy catalogue: 1) we use the halo catalogues of the \smdp\ simulation; 2) stellar masses are assigned using the \vmax-\ms\ relation assuming a total scatter of $\sigma = 0.15$ dex using the \citet{Rodriguez-Puebla+2020} GSMF; 3) an early- or late-type morphology is assigned based on the observed abundance in the SDSS DR7 galaxy groups catalogues \citep{Yang+2012} and the \citet{Huertas-Company+2011} morphologies; 4) \HI\ masses are assigned depending on whether the galaxy is central or satellite and whether the galaxy is central or satellite based on the \citet{Calette+2021} PDFs, and 5) our \HI-halo connection is totally defined. Selection effects or biases that emulate those of the observational surveys can be introduced in the mock catalogue.
} 
\label{fig:diagram} 
\end{figure*}

\section{The Galaxy-(Sub)Halo connection}
\label{methodology}

This section describes our procedure to attain the galaxy-(sub)halo connection by including both the empirical stellar and \HI\ mass distributions, and to obtain also predictions on the spatial clustering as a function \HI\ mass. The procedure implies the construction of a mock galaxy catalogue based on the outcome of a large cosmological N-body simulation. 
Figure \ref{fig:diagram} presents an illustrative scheme 
of all the ingredients of our procedure, and  it refers to 
the subsections where the details can be consulted. In short, we begin
by specifying the halo catalogue of the $N$-body simulation that 
we employed, Section \ref{sec:simulation}. Then, we describe 
how we assign stellar masses to every halo and subhalo in 
the simulation, Section \ref{sec:SHAM}, and their corresponding
late- or early-type morphologies, Section \ref{sec:Morphologies}.
Once we have specified the above properties for our mock galaxies, in Section \ref{sec:HI_mass}
we assign the \HI\ mass from empirical distributions for central and satellite galaxies.
Finally, every (sub)halo in the simulation ends with an assigned stellar and  \HI\ mass. As a result, the full distribution of \mha\ as a function of the (sub)halo circular velocity or mass is obtained and the \HI\ 2PCF can be measured. We can also apply to the mock catalogue cuts to emulate selection effects in the observational surveys for comparison (Section \ref{sec:selection_effects}).

\subsection{The simulation}
\label{sec:simulation}

We use the snapshot $z=0$ of the cosmological $N-$body \texttt{SmallMultiDark-Planck} (\smdp) 
simulation \citep{Klypin+2016} for (sub)halo properties and the spatial distribution of galaxies. 
The \smdp\ is a simulation with $3840^3$ particles and mass resolution 
of $9.63\times 10^{7} \msun/h$ in a box of $L=400h^{-1}$ Mpc on
a side. The adopted cosmology 
is based on the 
flat $\Lambda$CDM model with $\Omega_{\rm \Lambda}=0.693$, $\Omega_{\rm M}=0.307$, $\Omega_{\rm B}=0.048$, $h=0.678$, and $\sigma_8=0.829$, compatible with {\it Planck}15 results \citep{Planck+2016}.
Dark matter (sub)haloes were identified using the \rockstar\ 
halo finder \citep{Behroozi+2013a}. Masses or \vm\ for distinct haloes were defined using 
spherical overdensities according to the redshift-dependent virial overdensity  
$\Delta_{\rm vir}(z)$ given by the spherical collapse model; for the selected cosmology,  $\Delta_{\rm vir}= 333$ at $z = 0$. For subhaloes,
which is haloes that survived as bound entities inside the distinct haloes,
we used the highest mass or $\vm$ reached along the main progenitor branch of the subhaloes, see below. The (sub)halo catalogues are entirely downloadable\footnote{\url{https://www.cosmosim.org}} and their statistical properties were presented in detail in \citet{Rodriguez-Puebla+2016}.

\begin{figure*}
\includegraphics[trim = 0mm 40mm 50mm 10mm, clip, width=0.85\textwidth]{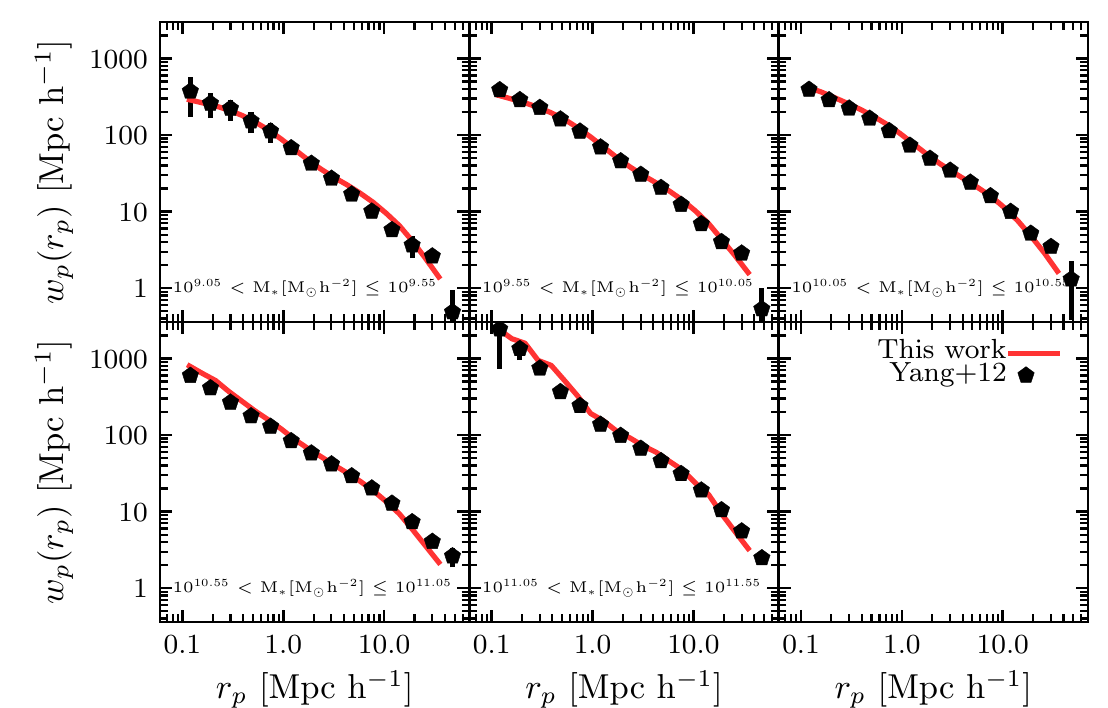}
\caption{Stellar 2PCF in different stellar mass bins measured directly from our mock galaxy catalogue. The red solid lines are our results and black pentagons are \citet{Yang+2012} inferences from SDSS observations for a given stellar mass bin, indicated at the top of each panel. }
\label{fig:stars-2PCFs} 
\end{figure*}

\subsection{Stellar masses}
\label{sec:SHAM}

For every halo and subhalo in the \smdp\ simulation a galaxy stellar mass is assigned via a \textit{non-parametric} SHAM technique. 
SHAM is a simple statistical approach for connecting a (sub)halo 
property (\mh{}, \vm, etc.) 
to that of a galaxy property (luminosity, \ms, etc.). In its simple form, SHAM assumes that
there is a unique monotonic relation between the selected (sub)halo 
and galaxy properties. 
The above clearly implies that central and satellite galaxies have identical galaxy-(sub)halo 
relationships. In the past, several authors have tested the effects of SHAM assumptions 
on the resulting galaxy clustering
\citep[see e.g.,][]{Yang+2009,Neistein+2011,Rodriguez-Puebla+2012,Rodriguez-Puebla+2013,Campbell+2017}. 
These studies have shown that using identical relationships between centrals
and satellites leads to inconsistent results as compared to  
the observed spatial clustering of galaxies. In particular,
\citet[][see also \citealp{Rodriguez-Puebla+2013}]{Rodriguez-Puebla+2012}
suggested that these relationships should be different in order
to not only reproduce the observed clustering of galaxies but also 
the observed abundance of galaxies in groups of different halo masses. 
Nonetheless, other studies have found that the assumptions from
SHAM can still be valid when instead of (sub)halo  
mass is used, for instance, \vm\ or the highest \vm\ reached along the main progenitor branch of the (sub)halo's 
merger tree, \vpeak\ \citep[][and more references therein]{Reddick+2013,Hearin+2013,Campbell+2017,Dragomir+2018,Wechsler+2018}. 
Following \citet[][]{Dragomir+2018}, here we assume that the halo property 
that correlates best with \ms\ and reproduces the observed galaxy spatial clustering is
\begin{equation}
	\vmax \equiv \left\{ 
			\begin{array}{c l}
				\vm & \mbox{for distinct haloes,}\\
				\vpeak & \mbox{for subhaloes.}
			\end{array}\right.
\label{vmax-def}
\end{equation}

Typically, authors use SHAM in its simplest form, that is, 
the cumulative halo and subhalo velocity function and the
cumulative GSMF are matched in order to determine the 
\ms-\vmax\ relation. The above assumes that there is
no scatter around the mean \ms-\vmax\ relation. 
In reality, it is expected that this relation 
has some scatter since the properties of the galaxies might be determined by different 
(sub)halo properties and/or some environmental factors. For example, in the case of the \ms--\mh\ relation, 
analysis of large galaxy group catalogues \citep{Yang+2009,Reddick+2013}, 
the kinematics of satellite galaxies  \citep{More+2011}, galaxy clustering
\citep{Shankar+2014,Rodriguez-Puebla+2015}, as well as galaxy clustering 
combined with galaxy lensing \citep{Zu+2015}, it is found 
that this dispersion
is of the order of $\sigma \sim 0.15-0.20$ dex.
To take this 
into account, SHAM should be modified to include a  
dispersion around the \ms-\vmax\ relationship. Thus, in this paper we use a more
general procedure for SHAM as described below. 

We define $\mathcal{H}(\ms|\vmax)$ as the conditional probability distribution function
that a halo with velocity in the range $\log\vmax\pm d\log\vmax/2$ hosts a galaxy 
with stellar mass in the range $\log\ms\pm d\log\ms/2$. We assume that
$\mathcal{H}(\ms|\vmax)$ is a lognormal function, 
\begin{equation}
\mathcal{H}(\ms|\vmax)=\frac{1}{\sqrt{2\pi\sigma^2}}\exp\left[-\frac{\left(\log\ms -\mu_M\right)^2}{2\sigma^2}\right],
    \label{ec:functional_H}
\end{equation}
where $\mu_M=\langle\log\ms(\vmax)\rangle$ is the mean logarithmic stellar mass-to-\vmax\ relationship and $\sigma$ is the scatter, assumed to be independent of \vmax.\footnote{Along this paper $\log$ is for logarithm base 10, and the width of the lognormal distribution, $\sigma$, is in dex. The calligraphic face is used for this case, that is, $\mathcal{H}(x|y) = H(x|y)\times\frac{x}{\log e}$.} 
Then the connection between the stellar mass and halo \vmax\ is given by the following equation:
\begin{equation}
    \phi_{\rm g}(\ms) =   \int \mathcal{H}(\ms|\vmax)\phi_{\rm DM}(\vmax)d\log \vmax,
    \label{ec:SHAM}
\end{equation}
where
\begin{equation}
    \phi_{\rm DM} (\vmax) = \phi_{\rm h}(\vm)  + \phi_{\rm sub}(\vpeak)
    \label{Eq:phi-vm}
\end{equation}
is the (sub)halo \vmax\ function and the units for $\phi_{\rm g}, \phi_{\rm DM}, \phi_{\rm h}$ and $\phi_{\rm sub}$ 
are in Mpc$^{-3}$dex$^{-1}$.

In this paper we use the GSMF, $\phi_{\rm g}(\ms)$, reported in Paper II, 
which ranges from dwarf galaxies, $\ms\sim 10^7\msun$, to massive galaxies at the centres of galaxy clusters, $\ms\sim 3\times10^{12}$. We note that the GSMF from Paper II 
was corrected for surface brightness incompleteness. 
For the total scatter around the \ms-\vmax\ relationship,\footnote{Note that this scatter is composed by the physical (intrinsic) scatter and random errors in the stellar mass determination. Estimates of random errors are of the order of  0.1 dex \citep[e.g.,][Paper II]{Tinker+2017}. The above implies 
that our assumed intrinsic scatter is then $0.11$ dex.} 
we assume $\sigma = 0.15$ dex in order to be consistent with the results discussed above for the \ms--\mhp\ relation (we notice that the scatter around the \ms--\vmax\ relation is expected to be lower than around the \ms--\mhp\ relation). 
We derive a non-parametric mean \ms-\vmax\ relationship by numerically deconvolving Equation (\ref{ec:SHAM}) as described in appendix D from Paper II.

The second panel of Figure \ref{fig:diagram} shows our obtained SHAM relationship between \ms\ and \vmax; see also Section \ref{results} below. Similarly to previous works, our relationship increases steeply with
\vmax\ for velocities below $\vmax\sim160$ km/s and it is shallower at higher velocities.
Finally, Figure \ref{fig:stars-2PCFs} shows that, as expected, our resulting mock galaxy catalogue reproduces well the observed SDSS DR7 projected 2PCFs at different stellar mass bins from \citet{Yang+2012}. For the 2PCFs, we assumed the plane parallel approximation with an observer along the $z$ direction. This is estimated by means of the \citet{LandySzalay1993} estimator and we integrate over the line of sight from $r_{\pi} = 0$ to $r_{\pi} = 40$ $h^{-1}$Mpc.

\subsection{Morphologies}
\label{sec:Morphologies}

In this subsection, we describe how we assign early and late-type morphologies to the galaxies in our mock catalogue. Notice, however, we assume that the scatter around the \ms--\vmax\ relation is independent of morphology, that is, the $P(\ms|\vmax)$ distribution is assumed to be the same regardless of morphology. 
The reason for this is that in the literature there is not yet a consensus on whether the galaxy-halo connection depends
on galaxy properties such as morphology, colour, or star formation rate. For example, 
some works find that at a fixed {\it halo mass} the stellar masses of blue/star-forming 
galaxies are on average larger \citep[e.g.,][]{More+2011,Rodriguez-Puebla+2015} than red/quenched 
galaxies, while other authors have found opposite results \citep[e.g.,][]{Moster+2018}. Others
simply do not find significant differences \citep{Zu+2016,Behroozi+2019}. Nonetheless,
all of them agree that at fixed {\it stellar mass}, red/quenched galaxies are more likely to reside in more massive halos than blue/star-forming galaxies,
consistent with weak lensing studies \citep[e.g.,][]{Mandelbaum+2006,Mandelbaum+2016}.
Since colour/star formation rate correlate with galaxy morphology, similar 
trends are expected when morphology is used instead. 
In any case, assuming that $P(\ms|\vmax)$ is independent of galaxy properties like morphology is
actually {\it irrelevant} for our results since we are interested here in predictions of the \HI-stellar-(sub)halo connection for the  \textit{whole} galaxy population. However, the introduction of
galaxy morphology is just a necessary step in order to assign \HI\ masses in our mock catalogue because the empirical \mha\ distributions as a function of \ms\ that we use are given separated into LTGs and ETGs.

Morphologies are assigned based on the observational fractions of ETGs for centrals, $f_{E}^{\rm cen}$, 
and satellites, $f_{E}^{\rm sat}$, as a function of stellar mass.  To do so, we first generate a
random number $U_{\mathcal{T}}$ uniformly distributed between 0 and 1. Then, for a given galaxy with
mass $\ms$, the morphology $\mathcal{T}$ is assigned as
\begin{equation}
	\mathcal{T} = \left\{ 
			\begin{array}{c l}
				{\rm ETG} & \mbox{if } U_{\mathcal{T}}< f_{E}^{i}(\ms), \\
				{\rm LTG} & \mbox{otherwise.}
			\end{array}\right.
\label{ec:morphology}
\end{equation}
The index $i$ refers either to a central or satellite galaxy. The fractions of ETGs for centrals and satellites were calculated from SDSS DR7 in \citet[][]{Calette+2021}, based on the galaxy group catalogue from \citet{Yang+2012}, the photometric catalogues of \citet{Meert+2015} and \citet{Meert+2016}, morphologies from \citet{Huertas-Company+2011}, and the volume corrections from \citet{Rodriguez-Puebla+2020}.

\subsection{HI masses}
\label{sec:HI_mass}

At this point, every halo and subhalo in the \smdp\ simulation at the snapshot $z=0$ has been assigned 
a stellar mass \ms\ and a galaxy morphology $\mathcal{T}$ depending on whether the galaxy is central 
or satellite. We use now these two galaxy properties to assign \HI\ masses. 

\begin{figure}
\centering
\includegraphics[trim = 5mm 43mm 57mm 12mm, clip, width=\columnwidth,height=190pt]{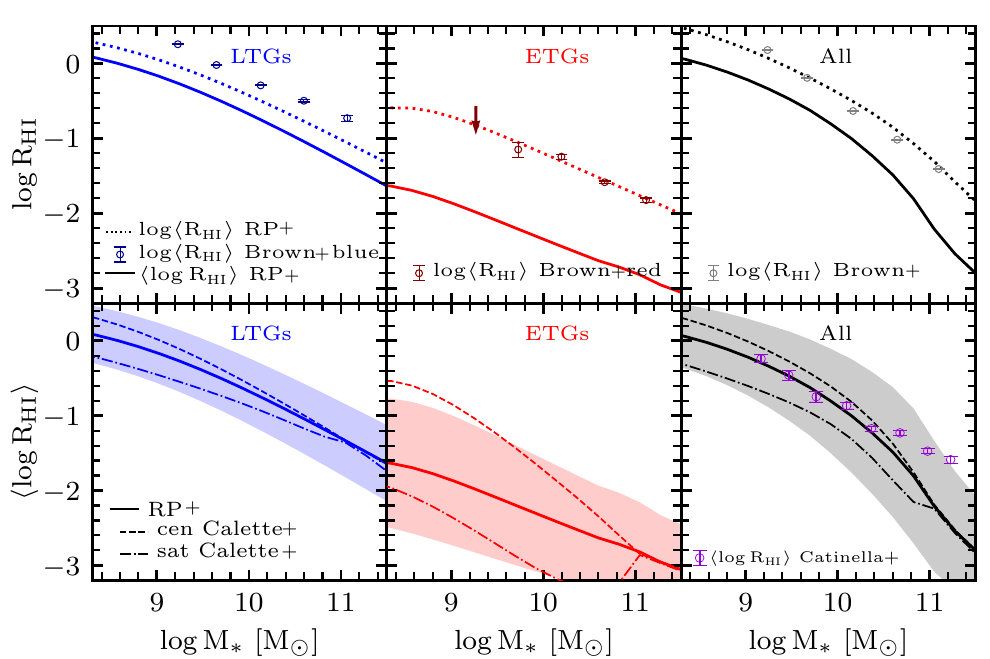} %
\caption{{\it Upper panels:} Empirical arithmetic and logarithmic mean \RHI--\ms\ relations (dotted and solid lines, respectively) computed from the \RHI\ conditional distributions reported in Paper II for LTG, ETGs, and all galaxies, from left to right, respectively. For comparison, we also plot the determinations by \citet[][]{Brown+2015}, who used \HI\ spectral stacking in bins of \ms\ and colour of a cross-match of the ALFALFA and SDSS surveys (open circles with error bars). {\it Lower panels:} Empirical logarithmic mean \RHI--\ms\
relations and their scatters (solid lines, as in the upper panels, surrounded by shaded regions) for LTGs, ETGs, and all galaxies, from left to right, respectively.
The corresponding logarithmic mean \RHI--\ms\ relations of central and satellite galaxies, calculated after applying the \texttt{xGASS}-based corrections described in the text and in \citet[][]{Calette+2021}, are shown in each panel with dashed and dot-dashed lines. The results of \citet[][]{Catinella+2018} from \texttt{xGASS} for all galaxies are plotted as open squares with error bars (errors of the mean); the \HI\ mass of non-detections were set to their upper limit values.
} 
\label{fig:Cen-sat-RHI_Ms} 
\end{figure}

In Paper I we found that the \mha\ (or $\RHI\equiv\mha/\ms$) conditional PDFs given \ms\ for LTGs are well described by a Schechter function while ETGs are well described by a Schechter 
function plus a top-hat function. The function parameters
for the \RHI\ distributions reported in Paper I were slightly improved in Paper II.
As mentioned in the introduction, our determinations homogenize and integrate the observational data from various local galaxy samples,
and, when necessary, allow us to estimate the \RHI\ distributions to values lower than the \HI\ flux limits of some samples. From these distributions, we may calculate the logarithmic $\langle\log\RHI\rangle-\log\ms$ relation and its scatter for LTGs, ETGs, and all galaxies, see solid lines surrounded by shaded bands in Figure \ref{fig:Cen-sat-RHI_Ms}.  In general, our $\langle\log\RHI\rangle-\log\ms$ relation for {\it all} galaxies is consistent with previous, more limited, determinations (see Paper I and \citealp{Calette+2021} for some comparisons; for the LTG and ETG relations there are not other works with which to compare). For example, the right lower panel of Figure  \ref{fig:Cen-sat-RHI_Ms} shows the $\langle\log\RHI\rangle-\log\ms$ relation as presented in \citet[][]{Catinella+2018} for the \texttt{xGASS} survey.
The largest differences are found at the high-mass end, where ETGs dominate. This is due to the fact that in \citet[][]{Catinella+2018}, as in other previous studies, the \HI\ mass of non-detections 
were set to their upper limit values, which leads to overestimating the means. 

In other studies, the authors report arithmetic means instead of logarithmic ones or use the technique of stacking the \HI\ spectra to increase the signal-to-noise ratio of non-detections; the stacking technique is equivalent to average arithmetically. As is well known, for distributions that are asymmetric, with a populated distribution on their low value side (as is the case with ETGs), these low values contribute little to the arithmetic mean in such a way that this mean is larger than the logarithmic one. From our full \RHI\ conditional PDFs, we can calculate also the arithmetic mean \RHI--\ms\ relations. By comparing our $\log\langle\RHI\rangle-\log\ms$ relations (dotted lines in the upper panels of Figure \ref{fig:Cen-sat-RHI_Ms}) with those presented in \citet[][]{Brown+2015} we find a reasonable agreement, as seen in Figure \ref{fig:Cen-sat-RHI_Ms}. These authors cross-match the ALFALFA and SDSS surveys and use the \HI\ spectral stacking technique in \ms\ and colour bins. Their \RHI--\ms\ relation for red galaxies agrees very well with our arithmetic mean \RHI--\ms\ relation for ETGs, showing that the arithmetic means are highly skewed towards the high values of \RHI, which correspond to detections. Since ALFALFA is biased towards \HI\ gas-rich and blue galaxies, it is expected that the \RHI--\ms\ relation for blue galaxies of \citet[][]{Brown+2015} should be above our relation for LTGs (note also that colour is not the same as morphology).

The \RHI\ conditional PDFs for LTGs and ETGs presented in Papers I and II do {\it not}
separate between centrals and satellites. Sampling \HI\ masses 
without taking into account centrals and satellites separately  could be problematic as it will be equivalent to ignoring any environmental processes in satellite galaxies, such as strangulation and gas stripping. At the same time, the \HI\ 2PCFs will be 
overestimated, specially at small distances, that is, at the 1-halo term. 
To overcome this problem, here we use the prescription presented in \citet{Calette+2021}
to obtain the \RHI\ conditional PDFs for LTGs and ETGs separated into central and
satellite galaxies.

Briefly, in \citet{Calette+2021} we used the \texttt{xGASS} survey \citep[][]{Catinella+2018}, which is a homogeneous sample that extends \texttt{GASS} \citep{Catinella+2010,Catinella+2013} down to $\ms\sim 10^9$ \msun{}. The \texttt{xGASS} detection limits in \mha\ are such that $\RHI>0.015$ for $\ms>3\times 10^{10}$ \msun\ and increases as \ms\ decreases up to $\RHI>0.02$ for $\ms=10^9$ \msun.
For non-detected galaxies in \HI, the upper limits of \mha\ are reported.
Galaxies are classified as satellites or centrals using the SDSS DR7 Group \citet{Yang+2007} catalogue and corrected for shredding \citep{Janowiecki+2017}. 
To assign the morphology $\mathcal{T}$, we use the \citet[][]{Huertas-Company+2011}
classification, mentioned above. In \citet{Calette+2021} we followed Paper I and applied to the \texttt{xGASS} sample the upper-limit corrections and the statistical procedures described there. Thus, we obtained the \RHI\ conditional distributions given \ms\ for LTGs and ETGs, {\it separated into centrals and satellites}. 
Then, we applied a continuous and joint fit to all these distributions;  for more details we refer the reader to \citet{Calette+2021}.

The fits to the \RHI\ distributions for all LTGs and ETGs from \texttt{xGASS}, which is not separating into centrals and satellites, are roughly consistent with those from Paper I.  However, there are three reasons why we cannot {\it directly} use the \mha\ distributions determined from  \texttt{xGASS} only to separate into centrals and satellites:
(1) \texttt{xGASS} extends down only to $\ms\sim 10^9$ \msun, (2) the \texttt{GASS} survey
is just a subsample of all the compiled data set in Paper I,
and (3) we want to be consistent with
the distributions reported in Paper I and updated in Paper II.  
To consider the above and at the same time holding the differences in the \RHI\ distributions of central and satellite found 
from \texttt{xGASS}, \citet{Calette+2021} 
computed the \RHI=\mha/\ms\ distributions separated into centrals and satellites as follows:
\begin{equation}
    \mathcal{P}_{i}^j(>\RHI|\ms)= \frac{\mathcal{P}_i^j(>\RHI|\ms)_{\rm xGASS}} {\mathcal{P}_i(>\RHI|\ms)_{\rm xGASS}} \times \mathcal{P}_i(>\RHI|\ms),
    \label{eq:corrections}
\end{equation}
where $i$ refers to either LTG or ETG,  $j$ to either central or satellite galaxy, and the sub-index \texttt{xGASS} refers to the respective distributions of \texttt{xGASS} galaxies as constrained in \citet{Calette+2021}. The distributions $\mathcal{P}_i$ correspond to the ones reported in Paper II.

The ratio of the  \RHI\ distributions from  \texttt{xGASS} in Eq. (\ref{eq:corrections}), 
should be considered as the factor needed to project the total \RHI\ distributions of LTGs and ETGs  into their corresponding distributions of centrals and satellite galaxies. The above warrants that our
total distributions are consistent with the ones reported in Paper II. 
Since \texttt{xGASS} extends only down to $\ms\sim 10^9$ \msun, \citet[][]{Calette+2021}  extrapolated to lower masses the fitted functions entering in the first term of the right side of Eq. (\ref{eq:corrections}), 
see for details that paper.
The fourth row panels of Figure \ref{fig:diagram} show what the \RHI\ conditional PDFs of centrals and satellites (both for LTGs and ETGs) look like as a function of stellar masses.
Figure \ref{fig:Cen-sat-RHI_Ms} shows the $\langle\log\mha\rangle-\log\ms$ relations and the standard deviations
around them as calculated from the distributions from Paper II
for LTGs, ETGs, and all galaxies (solid lines), as well as the corresponding mean relations for central and satellite galaxies after using Eq. (\ref{eq:corrections}) (dashed and dot-dashed lines, respectively). 

\begin{figure}
\includegraphics[trim = 5mm 70mm 107mm 12mm, clip, width=\columnwidth]{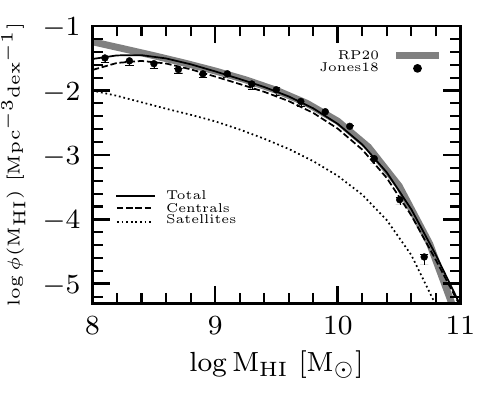}
\caption{\HI\ MF measured directly from our mock catalogue (solid line). The dashed and dotted lines correspond to centrals and satellites only, respectively. For comparison, we also show the \citet{Jones+2018} inferences (black circles) using ALFALFA 100\% ($\alpha.100$), and our empirical inference of the \textit{intrinsic} \HI\ MF from Paper II (thick solid line).  
} 
\label{fig:HIMF-all-cen-sat} 
\end{figure}

To assign \HI\ masses to our mock central or satellite galaxies, we first generate a
random number $U_{\rm HI}$ uniformly distributed between 0 and a 1. Then, for a given central or satellite galaxy with mass $\ms$ and morphology $\mathcal{T}$, the \HI\ mass is given by solving the following equation for \RHI,
\begin{equation}
	U_{\rm HI} =  \mathcal{P}_{i}^j(>\RHI|\ms),
\label{ec:HI_gass_mass}
\end{equation}
where $i$ refers to morphology $\mathcal{T}$ (ETG or LTG), 
and $j$ to either central or satellite galaxy. 
Since the \HI\ masses are assigned following robust empirical \RHI\ conditional PDFs given \ms\ and the
mock galaxy catalogue reproduces the observed GSMF by construction (see above), then one expects that the
catalogue should reproduce also the observed \HI\ MF (see Paper II).
Figure \ref{fig:HIMF-all-cen-sat} confirms this expectation. In this figure are 
also shown the \HI\ MFs of central and satellite galaxies separately. 
The former dominate by far the \HI\ gas content of galaxies at all masses. 
Notice that the downturn below $\mha\sim10^{8.5}\msun$ both for all and central galaxies 
is due to the \mha\ completeness limit in our mock catalogue, conditioned by the completeness limit of $\sim 10^7$ \msun\ in stellar mass (see Paper II).

Finally, our mock galaxy catalogue contains three properties both for central and satellite galaxies: stellar mass, \ms,  morphology, $\mathcal{T}$, and \HI\ mass, \mha. In other words, we have implicitly determined the multivariate conditional distribution function $\mathcal{P}(\mha,\mathcal{T},\ms|\vmax)$ that describes the connection between haloes and galaxies. Due to the assumptions used in this paper the above multivariate distribution is $\mathcal{P}(\mha,\mathcal{T},\ms|\vmax) = \mathcal{P}(\mha,\mathcal{T}|\ms) \mathcal{P}(\ms|\vmax)$.

\subsection{ALFALFA sample biases}
\label{sec:selection_effects}

The goal of this paper is to extend the galaxy-(sub)halo connection to the \HI\ component 
by means of the methodology described above. To check the robustness of our predicted results 
on the  \HI-(sub)halo connection it is key to compare the observed spatial clustering of galaxies as traced by their \HI\ mass with the respective clustering measured in our mock catalogue.

To date, the largest survey in \HI, where the projected 2PCF has been
measured for several \mha\ thresholds, is the ALFALFA radio survey \citep{Papastergis+2013,Guo+2017}.
However, a direct comparison between the resulting clustering of our mock galaxy catalogue and the ALFALFA survey is not trivial. To do so, we need  to consider in our mock catalogue the biases that the ALFALFA survey introduces to the galaxy population.
As discussed in the introduction, ALFALFA is a blind radio survey without any selection on target sample but its shallow $W_{50}$-dependent \HI\ flux detection limit introduces strong biases in some galaxy properties of the surveyed population other than \mha.

Here, in order to emulate the ``selection function'' of the ALFALFA survey, we use
the \RHI-\ms\ correlation reported in \citet[][see also \citealp{Huang+2012}]{Maddox+2015}. 
We assume that this correlation is lognormal distributed, and we interpolate the median and the dispersion
as a function of \ms\ as tabulated in their Table 1. We define this distribution as 
$\mathcal{S}(\RHI|\ms)$. Thus, for a given galaxy from our mock catalogue with properties $\ms$ and \mha, 
we generate a random number $U_S$ uniformly distributed 
between 0 and 1, and we determine whether 
the galaxy will be observed by the ALFALFA survey if
\begin{equation}
	\mathcal{S}(>\RHI|\ms) \geq U_S,
\end{equation}
otherwise it is not observed by ALFALFA. Note that the above
is {\it independent} of the galaxy morphology or whether the 
galaxy is central or satellite. As we will show below, applying the ALFALFA-like selection function will decrease the fraction of ETGs, as expected. This criterion will be applied to our catalogue when comparing 
the \HI\ spatial clustering with ALFALFA in \S\S \ref{secc:clustering}.

\section{Results}
\label{results}

\begin{figure*}
\includegraphics[trim = 10mm 45mm 50mm 5mm, clip, width=1.1\textwidth]{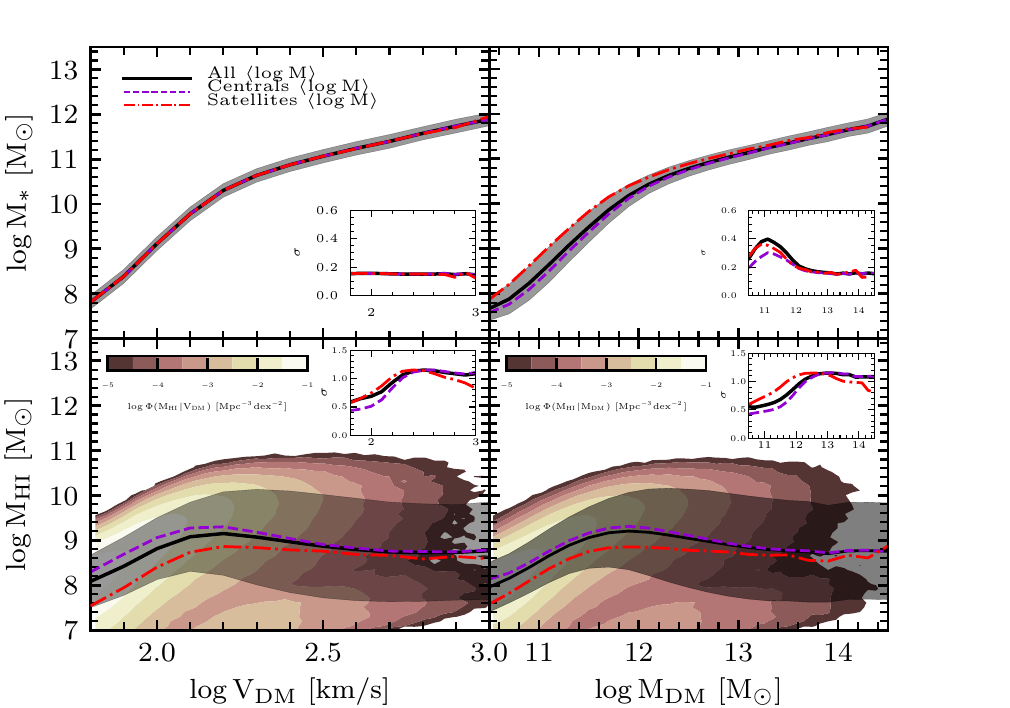}
\caption{Stellar and \HI\ galaxy-(sub)halo connection from the semi-empirical mock galaxy catalogue. \textit{Upper panels:} Mean logarithmic \ms--\vmax\ and \ms--\mh\ relations, solid black line, and their standard deviations, shaded region. The dashed and dashed-dotted lines correspond to the means of central and satellite galaxies. The insets show the scatter (SD) of each case. The scatter around the   \ms--\vmax\ relation was assumed the same for central and satellite galaxies and independent of any galaxy property.
\textit{Lower panels:} As the upper panels but for \mha. Notice that at difference of the \ms--\mhp\ relation, the \mha--\ms\ relation is not monotonic increasing and it has a huge scatter, specially at large masses. The coloured isocountours correspond to the bivariate (joint) distributions, $\Phi(\RHI|\vm)$ and $\Phi(\RHI|\mhp)$ in the left and right diagrams, respectively. Note that the distributions are bimodal, something inherited from the input \mha\ conditional distributions of LTGs and ETGs given \ms. 
} 
\label{fig:HI-halo-relations} 
\end{figure*}

Section \ref{methodology} describes how we generate a mock (sub)halo-galaxy catalogue in the \texttt{SMDP} N-body simulation. By construction the mock catalogue follows the observed total GSMF as well as the empirical \HI\ mass conditional PDFs of galaxies given their \ms. It is expected that the mock galaxy catalogue reproduces the observed
stellar projected 2PCFs and the \HI\ mass function. Indeed, in Figures \ref{fig:stars-2PCFs} and \ref{fig:HIMF-all-cen-sat} we showed that this is the case. 
Next, we present our results on the galaxy-(sub)halo connection including \HI\ mass
measured from our mock catalogue. 

In the upper panels of Figure \ref{fig:HI-halo-relations}, we show the mean $\langle\log\ms(\vmax)\rangle$ and $\langle\log\ms(\mhp)\rangle$ relations\footnote{Note that here \mhp\ is the virial mass \mh\ in the case of distinct haloes but for subhaloes this is the highest \mh\ along the main progenitor branch of their merger tree.}  
for all galaxies as well as the respective standard deviations.
Recall that the \ms--\vmax\ relation was assumed to be the same for central and satellite galaxies.  However, in the case of the \ms--\mhp\ relation (right upper panel) satellite galaxies segregate from centrals towards lower subhalo masses for a given stellar masses \citep[see e.g.,][]{Rodriguez-Puebla+2012}.
As was shown by these authors  \cite[see also][]{Rodriguez-Puebla+2013}, SHAM reproduces the observed spatial clustering of galaxies {\it only} when the stellar-to-(sub)halo relations of centrals and satellites are different in the direction shown in Fig. \ref{fig:HI-halo-relations}.
In our SHAM, this segregation emerges because \vmax\ is the halo property assumed to correlate better with \ms\ instead of \mhp, and, as discussed in \S\S \ref{sec:SHAM}, this form of SHAM is expected to generate results consistent with the spatial clustering of galaxies \citep[see also,][and \citealp{Dragomir+2018}]{Reddick+2013}.
Finally, note that the scatter around the \ms--\mhp\ relation increased with respect to the \ms--\vmax\ relation, specially at low masses, where the segregation between centrals and satellites is large.

Once we have shown that our mock galaxy catalogue is quite realistic, we can proceed to explore the predictions from this catalogue for \textit{(i)} the galaxy-(sub)halo connection for the \HI\ gas content (\S\S \ref{sec:Ext-Gal-Halo-Conn}), and \textit{(ii)}  the \HI\ projected 2PCFs for different \mha\ thresholds (\S\S \ref{secc:clustering}). These are the main goals of this paper.

\subsection{The galaxy HI-(sub)halo connection}
\label{sec:Ext-Gal-Halo-Conn}

The left and right lower panels of Figure \ref{fig:HI-halo-relations} show the full bivariate \RHI-\vm\ and \RHI-\mhp\ distributions, $\Phi(\RHI|\vm)$ and $\Phi(\RHI|\mhp)$, respectively. The colour isocountours cover a range of four orders of magnitudes. Note that the distributions at a given \vmax\ or \mhp\ are bimodal. This is inherited from the empirical \RHI\ conditional distribution given \ms. We also plot the logarithmic mean, $\langle\log\mha\rangle$, and its corresponding standard deviation for all galaxies (black solid line and shaded region) as well as the logarithmic mean for central (violet dashed line) and satellite (red dashed-dotted line) galaxies. The insets show the values of the respective standard deviations as a function of \vm\ and \mhp\ in dex. Tables \ref{table-main-VDM} and \ref{table-main-logMHI-logMD} in Appendix \ref{data-tables} provide the numerical values of, respectively, $\langle\log\mha(\vmax)\rangle$ and $\langle\log\mha(\mhp)\rangle$ along with their scatters.

On average, \mha\ for all galaxies increases roughly as $\mha\propto V_{\rm DM}^{3.5}$ or $\mha\propto\ M_{\rm DM}$ up to a peak at $\vmax\approx 160$ km/s or $\mhp\approx 10^{12}$ \msun, respectively; these trends are dominated by central galaxies, while satellite galaxies follow roughly similar trends but with lower values of \mha.  For large values of both \vmax\ and \mhp, the correlation becomes very weak. If any, $\langle\log\mha\rangle$ slightly decreases while the scatter strongly increases. The decrease of $\langle\log\mha\rangle$ is mainly due to centrals, while for satellites the mean relation keeps constant. At the largest velocities or masses, $\langle\log\mha\rangle$ is constant with \vmax{} or \mhp{}.  Our results show that {\it the \mha--\vmax\ and \mha--\mhp\ relations are neither monotonically increasing nor tight. } In \S\S \ref{discussion:SHAM}, we discuss the implications of the above for the traditional form of SHAM, and in \S\S \ref{discussion:comparisons} we compare our results with those from semi-analytical models and hydrodynamics simulations, some of which have predicted that the \mha--\mhp\ relation is non monotonic. In particular, the change of behavior of this relation at $\mh\sim10^{12}$ \msun\ is linked to the onset of AGN feedback \citep[][]{Kim+2017,Baugh+2019,Chauhan+2020}.  

The stellar-\HI--(sub)halo connection found here can be understood as the combination of the stellar mass conditional distribution given \vmax, $P(\ms|\vmax)$, and the empirical \HI\ mass conditional distribution given \ms, $P(\mha|\ms)$. The former was assumed to be {\it independent of galaxy morphology and environment}, while the latter strongly segregates by galaxy morphology and weakly by environment (defined by their central or satellite nature, see Fig. \ref{fig:Cen-sat-RHI_Ms} and \citealp{Calette+2021}). By analysing the two distributions mentioned above, we can interpret the results shown in Figure \ref{fig:HI-halo-relations}.
For $\vmax\lesssim 160$ km/s or $\mhp\lesssim 10^{12}$ \msun, the corresponding stellar masses are below $\ms\sim 10^{10}$ \msun. The galaxy population at these masses is completely dominated by LTGs, 
see Section \ref{secc:clustering} \citep[also Figure A1 in][]{Calette+2021}. For these galaxies, as seen in Figure \ref{fig:Cen-sat-RHI_Ms}, on average $\RHI\propto \ms^{-0.4}$, that is $\mha\propto \ms^{0.6}$. On the other hand, from Figure \ref{fig:HI-halo-relations}, we see that $\ms\propto V_{\rm DM}^{6.4}$ at low velocities. Therefore, roughly $\mha\propto V_{\rm DM}^{3.8}$, which is close to what it is seen in Fig. \ref{fig:HI-halo-relations} for $\vmax\lesssim 160$ km/s. For $\ms>10^{10}$ \msun, on average \RHI\ for all galaxies strongly decreases with \ms\ due to the increasing fraction of ETGs as \ms\ increases, which have much lower values of \RHI\ than LTGs at fixed \ms, see 
Fig. 5 in \citet[][]{Rodriguez-Puebla+2020}. The inferred dependence is $\RHI\propto \ms^{-1.2}$ or $\mha\propto \ms^{-0.2}$, combined with, $\ms\propto V_{\rm DM}^{1.7}$ for large velocities (Fig. \ref{fig:HI-halo-relations}), we expect $\mha\propto V_{\rm DM}^{-0.3}$, in rough agreement with what it is seen in Figure \ref{fig:HI-halo-relations} for $\vmax > 160$ km/s. For the largest velocities, the total \RHI--\ms\ relation flattens (see Fig. 5 in \citealp{Rodriguez-Puebla+2020}), hence the total \mha--\vmax\ relation flattens, too. 

The scatter distribution around the total \mha--\vmax\ relation is large. 
For $\vmax\lesssim 160$ km/s,  the standard deviations increase from $\sim 0.5$ to 0.9 dex. For central galaxies, the scatter is lower, while the opposite applies for satellites.
For larger velocities, the standard deviation increases up to 1.2 dex and at the largest \vmax\ values it slightly decreases. A large scatter is expected due to the broadness of the total \mha\ conditional PDFs at the stellar masses corresponding to these velocities. 
The total PDFs are broad because of (i) the strong difference between LTGs and ETGs in the distributions of \HI\ masses given \ms, see Fig. \ref{fig:Cen-sat-RHI_Ms}, and (ii) the large scatter around the mean \RHI--\ms\ relations of LTGs and ETGs, as seen in Fig. \ref{fig:Cen-sat-RHI_Ms}. An expected consequence of the former is that the scatter around the mean \mha--\vmax\ relation at the $\sim 160-400$ km/s range should be significantly segregated by morphology, with LTGs lying above the mean relation and ETGs below it.
In a forthcoming paper we will present these results for the LTG and ETG populations separately, and under different assumptions regarding the covariance of the scatter around the \vmax-\ms\ relation with morphology or colour. 
According to Fig. \ref{fig:HI-halo-relations}, around the \mha--\vmax\ relation there is also a (weak) segregation by central and satellite galaxies. Up to $\sim 400$ km/s, the former lies on average slightly above the mean \mha--\vmax\ relation and the latter below it. However, since this segregation is small, it is not expected to be a primary source of the large scatter around the \mha--\vmax\ relation.

\begin{figure}
\includegraphics[trim = 10.3mm 75mm 78mm 9mm, clip, width=\columnwidth]{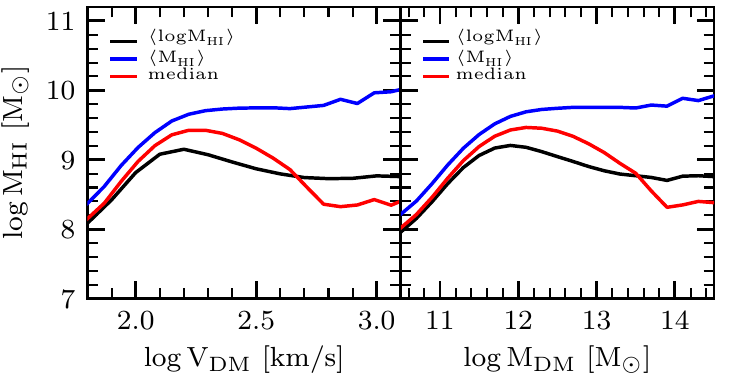}
\caption{The galaxy \HI-(sub)halo connection using three different statistical estimators: logarithmic mean, arithmetic mean, and median as labelled. The relations are different because the join distributions are asymmetric and even bimodal (see  Fig. \ref{fig:HI-halo-relations}). } 
\label{fig:HI-halo-stats} 
\end{figure}

It is worth mentioning that the scatter around the \mha--\vmax\ relation is not only large but also with a highly asymmetric distribution, and with even a bimodality at $\vmax>160$ km/s  as seen in Figure \ref{fig:HI-halo-relations}. As a result of the broad asymmetric distribution, different statistical estimators taken as representative of the population will significantly differ among them. 
Figure \ref{fig:HI-halo-stats} shows our predicted galaxy \HI-(sub)halo connection using the geometrical mean $\langle\log\mha\rangle$ (as in Fig. \ref{fig:HI-halo-relations}, black line), the arithmetic mean $\langle\mha\rangle$ (blue line), and the median $\overline{\mha}$ (red line). Up to $\vmax\sim 160$ km/s, where the scatter is relatively small and symmetric, the differences among the three statistical estimators are relatively small. 
For higher (sub)halo velocities, since the distribution of \mha{} presents a long tail towards low velocities, the arithmetic mean results much higher than the logarithmic mean and the median. The median differs from the geometric mean due to the increasing relevance at high velocities of the second peak in the \mha\ distribution at its low-velocity end. 

\begin{figure}
\includegraphics[trim = 9mm 74.9mm 45mm 8mm, clip, width=\textwidth]{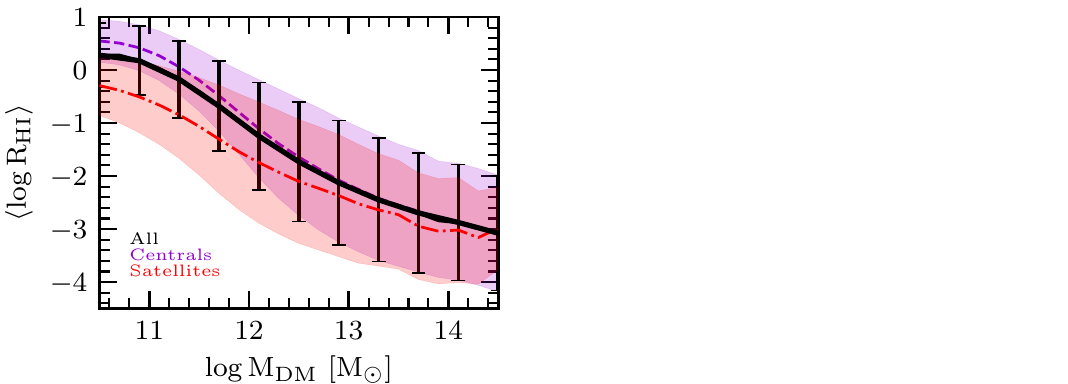}
\caption{Logarithmic mean of the \mha-to-\ms\ ratio as as function of (sub)halo mass, \mhp, for all galaxies (black solid line) as well as for only central (violet dashed line) and satellite (red dashed-dotted line) galaxies. The standard deviation of the relations for central and satellite galaxies are shown with the shaded regions.}  
\label{fig:RHI-Mh} 
\end{figure}

The explanations given above about the shape of the mean logarithmic \mha--\vmax\ relation and its scatter apply also to the shape of the mean logarithmic \mha--\mhp\ relation and its scatter (lower right panel of Fig. \ref{fig:HI-halo-relations}). 
The difference is that in the latter case, instead of $P(\ms|\vmax)$, we have $P(\ms|\mhp)$, which segregates by environment (see upper right panel of Fig. \ref{fig:HI-halo-relations}). The segregation of the \ms-\mhp\ relation between centrals and satellites produces a weaker segregation of the \mha--\mhp\ relation
between centrals and satellites than in the \mha--\vmax\ relation due to a compensation effect. Observations show that satellites have on average smaller \HI\ masses at a given \ms\ than centrals (see Fig. \ref{fig:Cen-sat-RHI_Ms}, and \citealp{Calette+2021}) but at a given \mhp, satellites have on average larger \ms\ than centrals. Hence, the difference in \mha\ at a given \mhp\ between satellites and centrals becomes smaller than in the case of a given \vmax. 
Distinct haloes contain on average more \HI\ gas than subhaloes of similar masses in the mass range of $\sim 10^{11}-10^{13}$ \msun, but the differences are small. 
In Figure \ref{fig:RHI-Mh} we plot the inferred \RHI--\mhp\ relation for all galaxies along together with the relations of central and satellite galaxies.
Note that the scatter around the mean \HI-to-halo mass relation of satellites galaxies is larger than for centrals, specially at $\mhp\lesssim 10^{12}$ \msun.

\subsubsection{The total galactic HI gas content inside distinct haloes}

\begin{figure}
\includegraphics[trim = 10mm 45mm 98mm 8mm, clip, width=\columnwidth]{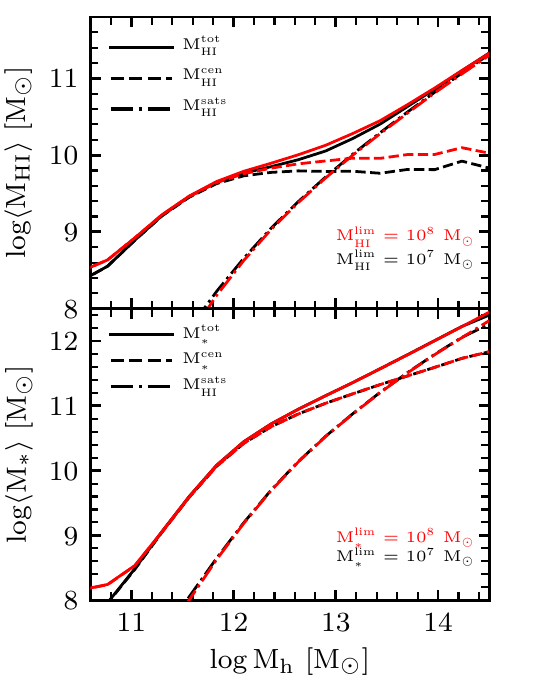}
\caption{Mean arithmetic \textit{total} \HI\ mass (top panel) and \textit{total} stellar mass (bottom panel) within a halo as a function of its virial mass (solid lines). The dashed and dot-dashed lines show the respective \HI\ and stellar masses contained only in the central galaxy and the sum of all satellites within the halo, respectively. Results for two threshold \HI/stellar masses are shown  (see labels in the panels).
}
\label{fig:logMHI-logMh-tot-cen-Sumsat} 
\end{figure}

Based on optical galaxy group catalogues for determining halo (group) masses, some authors have recently presented observational estimates of the \textit{total} \HI\ mass within these haloes, that is, the sum of \HI\ masses of the central and satellite galaxies in the halo, \mhatot = \mhacen + \mhaSsat\ \citep{Ai+2018,Obuljen+2019,Guo+2020,Tramonte+2020,Lu+2020}.  
The upper panel of Figure \ref{fig:logMHI-logMh-tot-cen-Sumsat} shows the mean $\langle\mhatot\rangle$ and standard deviation\footnote{Here we present arithmetic means because it is the way observations are presented \citep[see e.g.][]{Guo+2020} and we aim to compare with them.} as a function of \mh\ as measured from our mock catalogue. The mean for central galaxies, $\langle\mhacen\rangle$,  and the sum of all satellite \HI\ masses within a given distinct halo, $\langle\mhaSsat\rangle$,  are also plotted with dashed and dot-dashed lines, respectively. The two sets of lines are for two minimum \HI\  masses used to account for galaxies within the distinct halo, $M_{\rm HI,min}=10^7$ \msun, black lines, and $M_{\rm HI,min}= 10^8$, red lines. Columns (8)--(9) of Table \ref{table-main-logMHI-logMD} in Appendix \ref{data-tables} present the respective numerical values for the case $M_{\rm HI,min}=10^7$ \msun.

Figure \ref{fig:logMHI-logMh-tot-cen-Sumsat} shows that the dependence of \mhatot\ on \mh{} cannot be described by a double power law. In the $10^{12}\lesssim\mh/\msun\lesssim10^{13}$ regime the relation flattens. This is because for haloes up to $\sim 2\times 10^{12}$ \msun, central galaxies dominate by far the \HI\ gas content in the halos for which the relation flattens at these masses. For larger masses, the contribution of satellites increases and at $\mh\sim 10^{13}$ \msun\ the total \HI\ mass of satellites is already equal to the \HI\ mass of the central in the halo of same mass. At $\mh>10^{14}$ \msun\ the combined \HI\ gas of satellites is on average $\sim1$ dex larger than in the central galaxy. 
In the lower panel of  Figure \ref{fig:logMHI-logMh-tot-cen-Sumsat} we show the same as in the upper panel but for stellar mass. Two minimum stellar masses were used to sum up satellites, $10^7$ and $10^8$ \msun\ (the data corresponding to the former limit are presented in columns 10--11 of Table \ref{table-main-logMHI-logMD} in Appendix \ref{data-tables}). The differences between both cases are negligible. Unlike \mha, the dependence of \ms\ on \mh{} can be described by a double-power law.
This is mainly because the dependence of \ms\ on \mh{} for centrals does not flattens or decreases as it is the case for \mha.

\begin{figure}
\includegraphics[trim = 10mm 74mm 100mm 8mm, clip, width=\columnwidth]{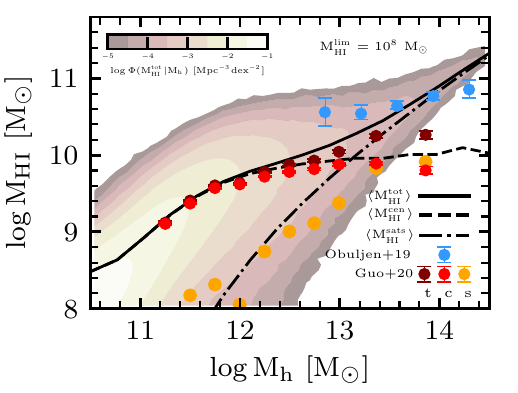}
\caption{Same as top panel of Figure \ref{fig:logMHI-logMh-tot-cen-Sumsat}, but now we compare our results for $M_{\rm HI}^{\rm lim}=10^{8}$ \msun\ with \citet{Guo+2020} and \citet{Obuljen+2019} observational inferences. Note that \citet{Guo+2020} inferred not only the total \HI\ content within haloes but also the contributions from the central. The isocountours correspond to the bivariate \mhatot\ and \mh\ distribution.}
\label{fig:logMHI-logMh-tot-cen-Sumsat-vs-obs} 
\end{figure}

We compare our semi-empirical inferences with previous ones in Figure \ref{fig:logMHI-logMh-tot-cen-Sumsat-vs-obs}. In this figure we show with isocountours the full bivariate \mhatot{} and \mh\ distribution. The total \HI\ mass estimates using stacked \HI\ spectra for galaxy groups by \citet[][]{Guo+2020} agree with our findings up to $\mh\sim 10^{13}$ \msun \ (we have transformed their halo masses defined at the $200\rho_m$ radius to our virial masses). For larger halo masses, their total \HI\ masses are lower than our determinations. As seen in Figure \ref{fig:logMHI-logMh-tot-cen-Sumsat-vs-obs} the above is mainly due to the lower total \HI\ mass they find for satellites, \mhaSsat, while the agreement with the \HI\ mass of centrals at all halo masses is encouraging. The low values of \mhaSsat\ may be due to the SDSS group catalogue not sampling the high halo mass end with enough statistics, due to the too low 21-cm line fluxes of many satellites as to contribute to the \HI\ stacked spectra above the allowed instrumental signal-to-noise ratio or due to the use of a too narrow window in the allowed \HI\ velocity in such a way that not all satellites are captured \citep{Chauhan+2021}.
On the other hand, the determinations for massive groups in \citet[][]{Obuljen+2019}, roughly agree with our results at the high halo mass range. Unlike \citet{Guo+2020}, \cite{Obuljen+2019} do not directly measure the \HI\ content of haloes, but instead use empirical relations to derive it. To determine \mhatot\ as a function of group (halo) mass they integrate the ALFALFA galaxy \HI\ MFs at different group masses. 

\subsection{The HI galaxy spatial clustering} 
\label{secc:clustering}

\begin{figure}
\includegraphics[trim = 3.2mm 73mm 112mm 10mm, clip, width=\columnwidth]{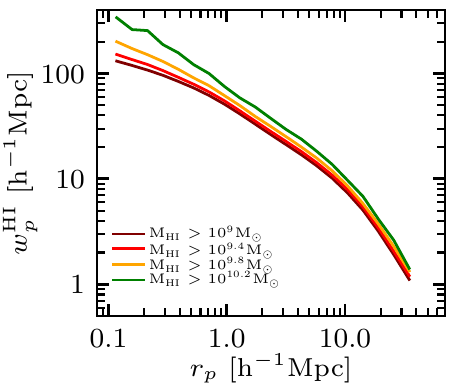}
\caption{Projected \HI\ 2PCFs from the galaxy mock catalogue measured above four \mha\ thresholds as indicated by the labels.}  
\label{fig:hi_2pcf_mock} 
\end{figure}

\begin{figure*}
\includegraphics[trim = 1.3mm 71mm 5mm 10mm, clip, width=1.05\textwidth]{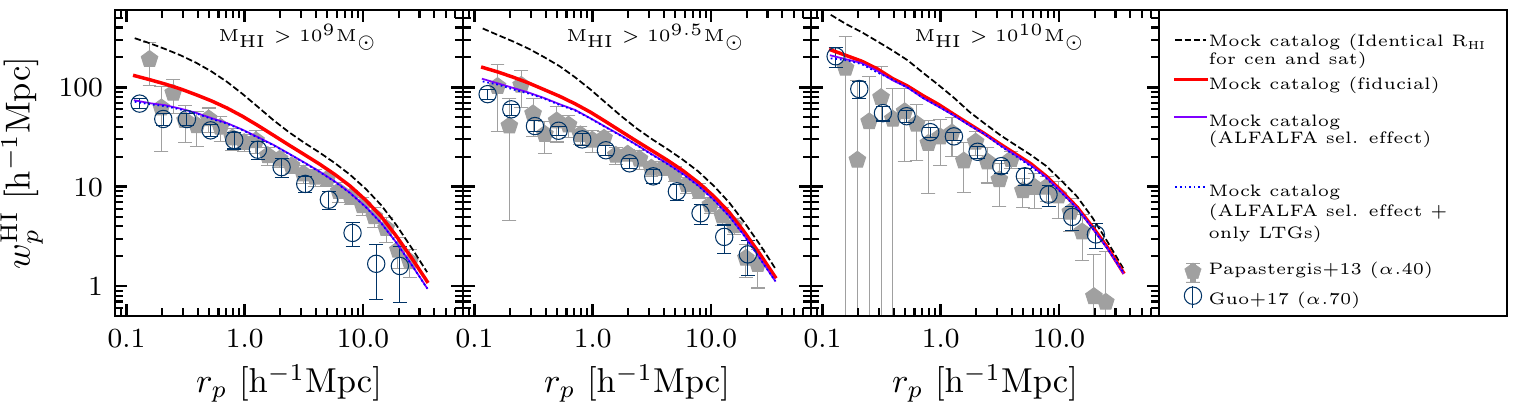}
\caption{Projected \HI\ 2PCFs measured from different \mha\ thresholds in the galaxy mock catalogue. The solid red line is our prediction for all local galaxies. This is the \HI\ spatial clustering expected from a large-volume and deep \HI\ blind survey. The black dashed line shows the projected 2PCFs from our approach but assuming the same \HI\ gas content for central and satellite galaxies. The violet solid line corresponds to the projected 2PCFs measured in the catalogue after emulating the ALFALFA survey selection effect (see text), while the blue dotted line is when the condition of not taking into account ETGs is also included. The symbols with error bars correspond to observational measurements from ALFALFA, see labels in the right-hand panel.} 
\label{fig:HI-2PCFs} 
\end{figure*}

We have discussed already that when implementing the SHAM for \vmax\ instead of \mhp, the projected 2PCFs, $\omega_p^{\rm *}(r)$, from the SDSS DR7 galaxies in stellar mass bins are recovered, see Figure  \ref{fig:stars-2PCFs}. 
Next, we present our predictions for the projected 2PCFs as a function of \HI\ mass.

Figure \ref{fig:hi_2pcf_mock} shows the resulting \HI\ projected
2PCFs, $\omega_p^{\rm HI}(r)$, for various \mha\ thresholds from our mock galaxy catalogue. In general, 
the amplitude of $\omega_p^{\rm HI}(r)$ is nearly independent of the
\mha\ threshold. In more detail, however,  we notice that the
amplitude of the one-halo term increases
with \mha\ while the two-halo term is almost independent of \mha, and in consequence, the resulting projected 2PCFS are not power laws. 
The above can be broadly understood as the result of the differences 
between the $\mha-\vmax$ (or $\mha-\mhp$) relations of centrals and satellites and the flattening of both relations with similar values of the mean $\langle\log\mha\rangle$
at high halo masses. Note that for low \mha\
thresholds central galaxies contribute more to $\omega_p^{\rm HI}(r)$ than satellites but for high \mha\ thresholds, centrals and satellites contribute equally to $\omega_p^{\rm HI}(r)$. 
This behaviour of 
$\omega_p^{\rm HI}(r)$ is very different to the one of
$\omega_p^{\rm *}(r)$, as 
$\omega_p^{\rm *}(r)$ depends strongly on stellar mass. 

There are other features related to the \HI\ projected 2PCFs that are worth mentioning. We begin by emphasizing the fact that the 2PCFs mass thresholds used
in Fig. \ref{fig:hi_2pcf_mock} are usually above the mean $\mha-\vmax$ and $\mha-\mhp$ relations, see Fig. \ref{fig:HI-halo-relations}. In other words, the larger the \HI\ mass threshold, the larger the distance from, for example, the $\langle\log\mha\rangle$ values.
The above is interesting for various reasons. First, increasing the \HI\ mass threshold implies that
we are sampling {\it increasingly rarer} masses for a given $\vmax$ or $\mhp$. 
Secondly, similar mass thresholds have been used in
the past to derive the \HI\ galaxy-halo connection \citep[e.g.,][]{Guo+2017}. Recall that HOD (and related) models assume that
galaxy properties, in this case \mha, are {\it totally} 
determined by a halo property such as $\vm$ or $\mh$. 
The lower panels of Figure \ref{fig:HI-halo-relations} show that
this is not the case. In view of the above, it is thus relevant to
ask whether HOD (and similar) models are appropriate 
to constrain the {\it real} \HI-to-halo mass relation. We will come back to this point in Section \ref{discussion}.

Figure \ref{fig:HI-2PCFs} shows again the \HI\ projected 2PCFs, $\omega_p^{\rm HI}(r)$, for different \mha\ thresholds as indicated in
the panels. The thick red lines correspond to the measurements from our mock catalogue.
One of the key aspects in our mock catalogue is that we allowed the \HI\ mass conditional distributions for a given \ms, both for LTGs and ETGs, {\it to be different for centrals and satellites} as observations show, see Section \ref{sec:HI_mass}.
The black dashed lines in Figure \ref{fig:HI-2PCFs} show the resulting $\omega_p^{\rm HI}(r)$ functions  
when assuming that the \HI\ distributions are the same for centrals and satellites,
that is, the ratio in Eq. (\ref{eq:corrections}) is set equal to 1. The fact that satellite galaxies have lower \HI\ gas fractions than centrals clearly works in the direction of lowering $\omega_p^{\rm HI}(r)$ at small distances, i.e. at the 1-halo term. 

In Figure \ref{fig:HI-2PCFs}, we compare
our results  to the corresponding $\omega_p^{\rm HI}(r)$ functions from the ALFALFA $\alpha.40$ survey \citep{Haynes+2011} as measured by \citet[][pentagons]{Papastergis+2013} and from the ALFALFA $\alpha.70$ survey as measured by  \citet[][open circles]{Guo+2017}. Beyond the differences between both studies, it is evident that our predicted \HI\ 2PCFs are above both studies. At this point, it is important to recall that while the ALFALFA survey is blind in radio, it actually suffers from biases in galaxy properties different to \mha, see introduction. The main reason for these  biases is related to the low detection limit in \HI\ flux and its dependence on the \HI\ line width \citep{Haynes+2011,Huang+2012,Chauhan+2019}. As a result,
the \RHI--\ms\ relation of ALFALFA is biased towards \HI-rich galaxies, avoiding galaxies with intermediate to low values of \RHI\ (\citealp{Huang+2012}; \citealp{Maddox+2015}; see Figure 6 of \citealp{Rodriguez-Puebla+2020}).
 \HI-poor galaxies tend to be earlier types and reside in denser environments, hence, they are expected to be more clustered than \HI-rich galaxies. Therefore, the observed low clustering of ALFALFA observations as compared to our predictions is most likely due to the lack of \HI-poor highly clustered galaxies in the ALFALFA survey.
 The $\omega_p^{\rm HI}(r)$ functions from our mock galaxy catalogue showed in Figure \ref{fig:HI-2PCFs}
 should be taken as predictions to compare with what will be measured in future deep and large \HI\ surveys, not significantly affected by sample biases.

Section \ref{sec:selection_effects} describes a simple way to emulate the non-trivial selection effects of the ALFALFA survey. The violet lines in Figure \ref{fig:HI-2PCFs} show
the resulting $\omega_p^{\rm HI}(r)$ functions from our mock catalogue after applying this ALFALFA-like selection. 
The functions approximate those reported from ALFALFA, specially for the low \mha\ threshold. Nonetheless, observations still show slightly lower amplitudes than our predictions.

\begin{figure}
\includegraphics[trim = 8mm 74mm 100mm 7mm, clip, width=1.1\columnwidth]{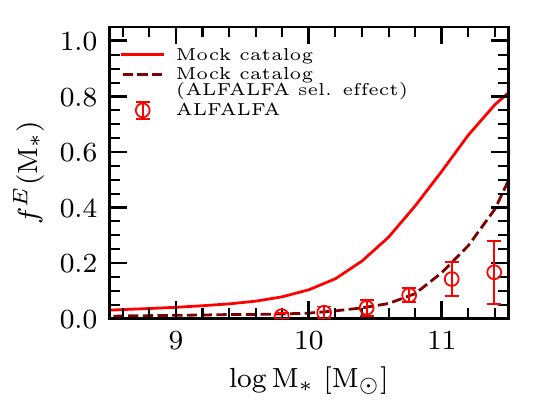}
\caption{Fraction of ETGs as a function of \ms\ as measured in the complete galaxy mock catalogue (solid line) and after imposing the ALFALFA-like selection (dashed line). The corresponding fractions measured in the ALFALFA survey in many stellar mass bins are showed with circles and error bars. 
} 
\label{fig:fE-alfalfa} 
\end{figure}

By imposing the  ALFALFA-like selection effect in our mock catalogue, the fraction of ETGs is decreased as expected, but when comparing our fraction of ETGs as a function of \ms\ with ALFALFA, the latter shows lower ETG fractions even at high masses. Figure \ref{fig:fE-alfalfa} shows the above comparison. 
This is because massive ETGs have both low \HI\ content and large $W_{50}$ equivalent widths both of which act against detection in this and other \HI\ blind surveys \citep[see][]{Obreschkow+2013,Chauhan+2019}. Moreover, the volume of ALFALFA is relatively small, so massive galaxies, typically ETGs, are underrepresented due to cosmic variance. In order to explore an extreme case,
in addition to the ALFALFA-like selection effect, we assume that ALFALFA observes LTGs only. The resulting measured  $\omega_p^{\rm HI}(r)$ are plotted in Figure \ref{fig:HI-2PCFs} with blue dotted lines. They are almost indistinguishable from the previous case. 

Our analysis shows that while the selection effects on properties different to \mha\ introduced by the shallowness of a blind \HI\ survey, like ALFALFA, is not crucial for measuring the \HI\ MF (our \HI\ MF is consistent with ALFALFA, see Fig. \ref{fig:HIMF-all-cen-sat}), these selection effects become critical for measuring the \HI\ spatial clustering (as well as for the \HI-to-stellar mass correlation, see Papers I and II). In \S\S \ref{sec:caveats} we discuss the caveats of our approach and how, by overcoming them, we could improve our comparison with the galaxy \HI\ clustering from ALFALFA.

\section{Discussion}
\label{discussion}

\subsection{ HI mass is not determined by the halo scale: implications for SHAM and HOD}
\label{discussion:SHAM} 

The assumption that the halo mass/maximum circular
velocity determines the 
properties of the galaxies is central to galaxy-halo models such as the HOD and SHAM. Previous works have shown that, this is a good approximation for galaxy properties 
such as luminosity and stellar mass \citep[for a discussion, and references see][]{Dragomir+2018}. As discussed in 
Section \ref{secc:clustering}, when using \mha\ 
the central assumption of  SHAM or HOD models is clearly
flawed as can be seen in the lower panels of Fig. \ref{fig:HI-halo-relations}. Thus expecting to derive 
a \mha--\vmax\ (or \mha--\mhp) relationship from these models under the assumptions above is fruitless. 
In particular, it is clear that, by construction, 
SHAM will tend to predict an increasingly 
monotonic mean relation between \mha\ and \vmax\ (or \mhp);
a relation that will fail to reproduce 
the observed clustering from \HI-surveys as discussed in \S\S \ref{secc:clustering}. Here, we
elaborate more on this argument by deriving some 
\mha--\vmax\ relations from SHAM and studying their resulting 
clustering in \HI\ mass.

\begin{figure}
\includegraphics[trim = 10mm 74mm 105mm 7mm, clip, width=1.1\columnwidth]{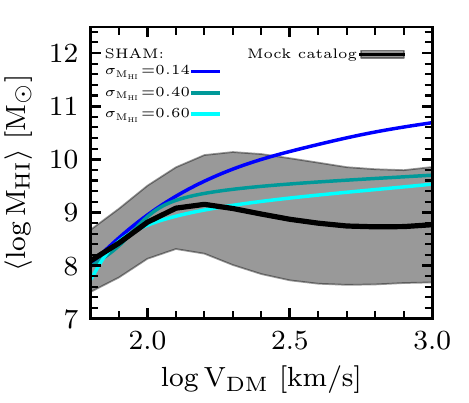}
\caption{SHAM results using the observed \HI\ MF for three different assumptions on the magnitude of the scatters for the \mha--\vmax\ relation (see labels). The black solid line and shaded area are our result as shown in Figure \ref{fig:HI-halo-relations}. } 
\label{fig:simple-AMT} 
\end{figure}

Figure \ref{fig:simple-AMT} presents three different \mha--\vmax\ relations obtained  by means of SHAM (see Section \ref{sec:SHAM}) using the empirical HI\ MF from \citet[][]{Rodriguez-Puebla+2020}.  
We assume that the scatter around the mean logarithmic \mha-\vmax\ relation is normally distributed and use three different values for the scatter: $\sigma=$0.15 (blue line), 0.40 (green line), and 0.60 (cyan line) dex. The black solid line and the shaded area reproduce our predicted mean \mha--\vmax\ relation and its scatter as showed in  Fig. \ref{fig:HI-halo-relations}. It is clear that SHAM predicts that the 
average relations increase monotonically with \vmax. Notice, however, that by increasing the scatter, the SHAM results approach our \mha--\vmax\ relation. Nonetheless, by construction, SHAM
will always introduce a monotonic correlation between \mha\ and \vmax\ 
or \mh\ regardless of the scatter assumed. 

\begin{figure*}
\includegraphics[trim = 1.3mm 71mm 5mm 10mm, clip, width=1.05\textwidth]{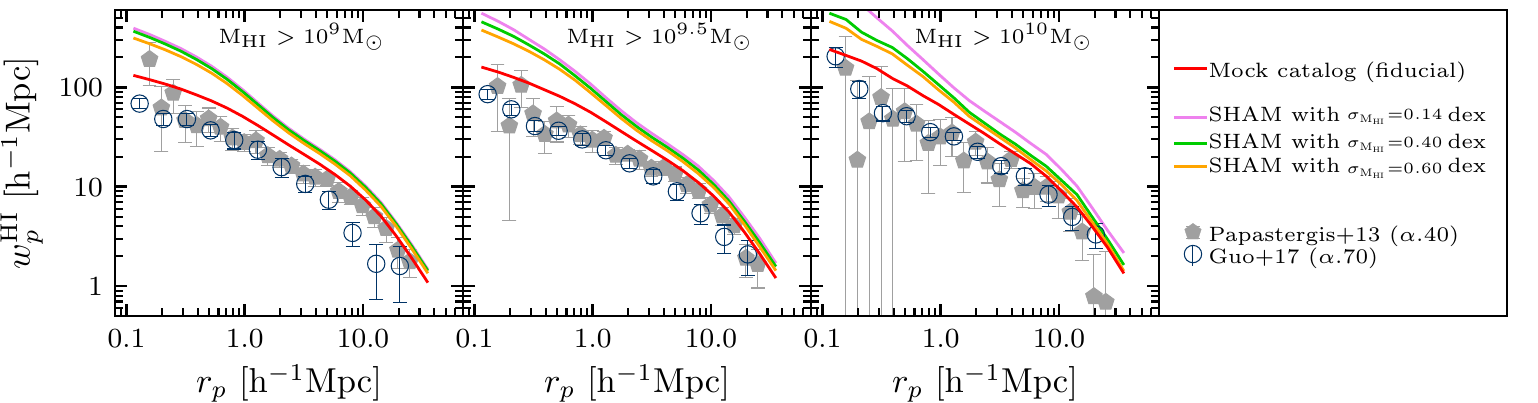}
\caption{As Figure \ref{fig:HI-2PCFs} but using the different SHAM models shown in Fig. \ref{fig:simple-AMT}. The solid line is as in Figure \ref{fig:HI-2PCFs} and it corresponds to the prediction from our galaxy \HI-(sub)halo connection. 
} 
\label{fig:HI-2PCFs-vs-am} 
\end{figure*}

Figure \ref{fig:HI-2PCFs-vs-am} shows the resulting
\HI\ clustering based on our SHAM experiments. 
We note that there are two potential flaws that are 
evident for SHAM. First, the fact that 
SHAM does not separate between centrals and satellites. 
As discussed in Section \ref{sec:HI_mass} this assumption is problematic
as it is similar to ignoring environmental effects for 
satellite galaxies, so it is not surprising that 
the one halo-term is overestimated. Secondly, 
since SHAM introduces a monotonic correlation between \mha\ and \vmax, 
the amplitude of the two-halo term will 
always increase with \mha,  thus overestimating the 
two-halo term. Note that as we decrease the scatter, the tighter and
stronger the correlation with \vmax\ is, and thus the larger 
the amplitude of the 2PCFs. A similar result has been reported in \citet[][]{Guo+2017} and \citet[][]{Stiskalek+2021}.

\begin{figure*}
\includegraphics[trim = 5mm 74.5mm 55mm 7mm, clip, width=\textwidth]{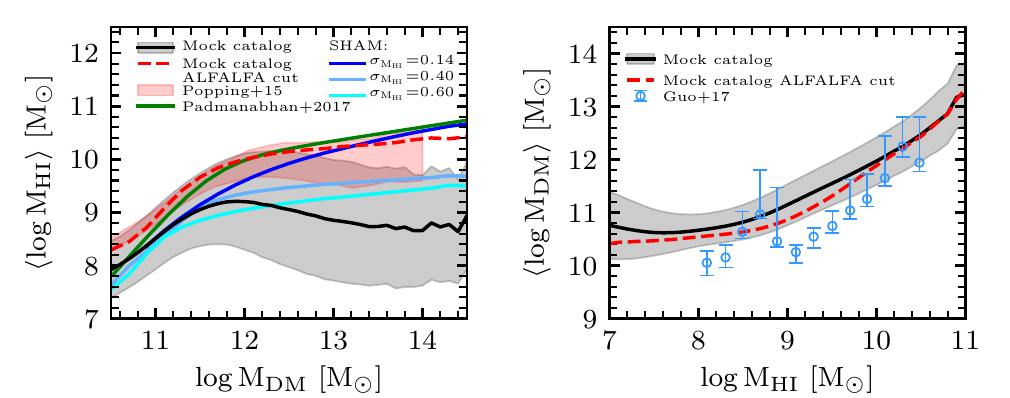}
\caption{ {\it Left-hand panel:} Logarithmic mean \mha--\mhp\ relation and its scatter as shown in Fig. \ref{fig:HI-halo-relations} (black solid line and gray shaded area) and the resulting relation when applying the ALFALFA-like selection criterion (dashed red line), which selects \HI-rich galaxies. The results from the SHAM models presented in Fig. \ref{fig:simple-AMT} are also shown here. The green solid line is for the SHAM model (without scatter) from \citet[][]{Padmanabhan+2017}, and the pink shaded area is for the model for star-forming galaxies presented in \citet{Popping+2015}. 
{\it Right-hand panel:} The inverse mean \mha--\mhp\ relation (it is almost indistinguishable for measurements of only central galaxies). Line symbols are as in the left-hand panel. The circles with error bars correspond to results from an SHAM model extended to introduce assembly bias and constrained with the ALFALFA \HI\ clustering \citep{Guo+2020}. }
\label{fig:HOD-comparisons} 
\end{figure*}

The left-hand panel of Figure \ref{fig:HOD-comparisons} is as Figure \ref{fig:simple-AMT}  but for the halo mass \mhp.
We used the tight relation between \vmax\ and \mh\ as measured in N-body simulations \citep[][]{Rodriguez-Puebla+2016} to convert the \mha-\vmax\ to \mha-\mh. We show in this plot a previous determination by \citet[][green solid line]{Padmanabhan_Kulkarni2017} of the \HI-to-halo mass relation with SHAM using the observed \HI\ Mass Function. The SHAM implemented by these authors did not include scatter; this is why their \HI-to-halo mass relation looks similar to our case with small scatter, blue line. As can be seen, this relation is far from what we actually found with our semi-empirical model.

In the past, some authors used in different ways the \HI\ spatial clustering information provided by current blind \HI\ surveys for constraining the galaxy \HI-halo connection by means of HOD or SHAM models  \citep[e.g.,][]{Padmanabhan+2017,Guo+2017,Obuljen+2019}. As discussed in \S\S \ref{secc:clustering}, the \HI\ projected 2PCFs are measured for galaxies with masses larger than a given \mha\ threshold, and the values of this threshold are typically above the mean \HI-to-halo mass relation as determined here, implying that the measured 2PCFs correspond to a \textit{biased} galaxy population.
On the other hand, it is well known that \HI\ blind surveys are strongly biased to gas-rich, blue galaxies (see introduction). Both shortcomings are related because the minimum \HI\ mass threshold is determined by the \HI\ mass completeness limit of the survey, and the latter is partially related to the \HI\ survey sensitivity, which introduces a bias against gas-poor, red galaxies (typically ETGs).

In Figure \ref{fig:HOD-comparisons} we show the \mha--\mhp{} relation (left-hand panel) and its inverse, the \mhp--\mha{} relation (right-hand panel), as measured from our catalogue when imposing the ALFALFA-like selection, red dashed lines. Recall that with this selection, the $\omega_p^{\rm HI}(r)$ function with a threshold of $\mha=10^9$ \msun\ resulted in a reasonably agreement with the ALFALFA measurements (Fig. \ref{fig:HI-2PCFs}).  
For comparison, the respective relations along with their standard deviations as measured from our galaxy mock catalogue without imposing any selection criterion, are also plotted in Figure \ref{fig:HOD-comparisons}. An implication of this figure and Fig. \ref{fig:HI-2PCFs} is that by (roughly) reproducing the ALFALFA \HI\ spatial clustering, the galaxy \HI-halo connection is biased to high \mha\ values at a given halo mass or to low halo masses at a given \mha. 
In the latter case, the difference would be larger if the ALFALFA \HI\ clustering would be better reproduced by assuming that the scatter around the \ms--\vmax\ relation depends on morphology (see \S\S \ref{sec:Caveats-clustering} below). 

In the right-hand panel of Figure \ref{fig:HOD-comparisons}, the results from the SHAM model of \citet[][]{Guo+2017} for central galaxies are reproduced. 
The inverse \mha--\mhp\ relation in our case refers to haloes and subhaloes (central and satellite galaxies); however, we note that for distinct haloes (centrals) the results remain similar.
\citet[][]{Guo+2017} found that their simple SHAM model predicts much higher \HI\ clustering than the measurements from ALFALFA. 
Then, they extended the model by introducing other halo parameters such as the halo formation time. 
When using the 2PCFs from ALFALFA to constrain their model, the authors found that \HI\ galaxies are more likely to reside in late-forming haloes (assembly bias).\footnote{ \citet[][]{Guo+2017} have also applied to their mock catalogue the $W_{50}$-dependent flux detection limit of ALFALFA but did not present their effects on the \HI\ clustering explicitly as we do. It is likely that these effects are outweighed by the formation-time halo selection.}
When applying the ALFALFA-like selection function to our mock catalogue, our \mhp--\mha\ relation (red dashed line) 
is more consistent with the one from the \citet{Guo+2017} model, although our relation is slightly above it. Recall that the \citet{Guo+2017} model matches the ALFALFA \HI\ clustering by construction. 
In section \ref{sec:Caveats-clustering}, we discuss potential modifications to our model that could help to reproduce better the ALFALFA \HI\ clustering. 

Finally, the left-hand panel of Figure \ref{fig:HOD-comparisons} 
presents the results based on the more physically motivated inferences of \HI\ mass from
\citet{Popping+2015}  for {\it star-forming} galaxies. 
In order to infer \HI\ gas mass, the authors used the star formation histories from SHAM and the inverted star formation rate–surface density relations to infer galaxy HI masses. It is interesting how our  \mha--\mhp\ relation for {\it gas-rich} galaxies (due to the imposed ALFALFA-like selection, dashed line)
is similar to the one of those authors. Notice that \citet{Popping+2015} results for \HI\ gas mass were not obtained from a SHAM but using a physically motivated model.

\subsection{Caveats and the galaxy HI clustering of ALFALFA}
\label{sec:caveats}  

\subsubsection{Morphological classification}

In this paper we use the observed fractions of ETG for 
centrals and satellites based on the automatic
morphological classification from \citet{Huertas-Company+2011}.
As discussed in Section \ref{sec:Morphologies}, we introduce
galaxy morphology just as a necessary step in order to assign
\HI\ masses to our galaxies. Thus, a potential concern is the 
effects of using alternative morphological classifications 
in our methodology. To study the above, we use 
an alternative morphological classification for SDSS galaxies.

\begin{figure}
\includegraphics[trim = 7mm 74mm 105mm 7mm, clip, width=1.1\columnwidth]{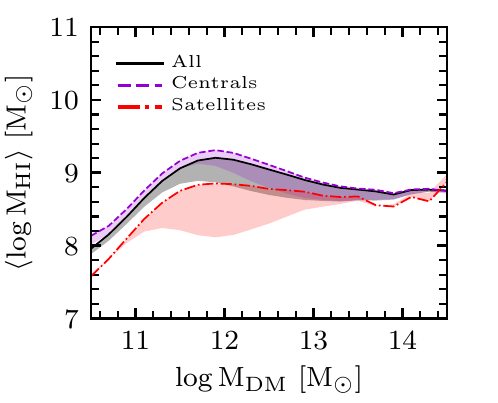}
\caption{The lines show the logarithmic mean \mha--\mhp\ relations for all (black), central (violet), and satellite (red) galaxies using the \citet{Huertas-Company+2011} morphological classification for defining LTGs and ETGs, as shown in Figure \ref{fig:HI-halo-relations}. 
The lower extremes of the shaded areas show the same relations but when using the \citet{Dominguez-Sanchez+2018} morphological classification. The latter relations are below the former ones, specially for satellites. The shaded area can be considered as the uncertainty in our method due to differences in the morphological classification.} 
\label{fig:HI-halo-relations-DS18} 
\end{figure}

\begin{figure}
\includegraphics[trim = 2mm 70mm 110mm 8mm, clip, width=1.05\columnwidth]{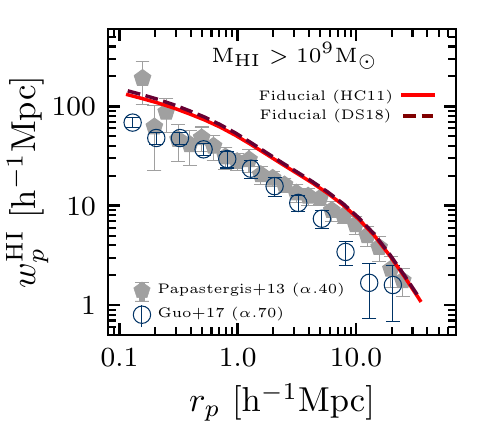}
\caption{Projected \HI\ 2PCF for $\mha>10^9$ \msun\ for our fiducial model using the \citet[][red solid line]{Huertas-Company+2011} and the \citet[][dashed line]{Dominguez-Sanchez+2018} morphological classifications. The observations are as in Figure \ref{fig:HI-2PCFs}} 
\label{fig:hi_2pcf_mock-DS19} 
\end{figure}

Based also on SDSS, \citet[][]{Dominguez-Sanchez+2018} presented a new automatic morphological classification of galaxies. As shown in \citet{Calette+2021},
the \citet{Dominguez-Sanchez+2018} classification implies a higher fraction of both central and satellite ETGs than \citet{Huertas-Company+2011} up to $\ms\sim 10^{11}$ \msun.
When we use the fractions from \citet[][]{Dominguez-Sanchez+2018} to assign \HI\ masses, we note that, as expected, some of the results for the \HI-halo connection change. 
Figure \ref{fig:HI-halo-relations-DS18} shows the logarithmic mean \mha--\mhp\ relation for all galaxies and for centrals and satellites using both the \citet{Huertas-Company+2011} (same line code as in Fig. \ref{fig:HI-halo-relations}) and \citet[][the lower extreme of the shades areas]{Dominguez-Sanchez+2018} morphological classifications. 
For the latter classification, the values of $\langle\log\mha\rangle$ are lower than for the former classification, in particular for satellite galaxies. This is expected because, as mentioned above,  \citet[][]{Dominguez-Sanchez+2018} classify a larger fraction of ETGs, specially in the case of satellites, than \citet{Huertas-Company+2011}, and ETGs contain less \HI\ than LTGs of the same stellar mass.  The shaded areas in this figure can be considered as systematic in our inferences due to the uncertainty in the galaxy morphological classification. Similarly, in Figure \ref{fig:hi_2pcf_mock-DS19}, we plot the \HI\ projected 2PCFs from our mock catalogue using both morphological classifications. Inferences using \cite{Huertas-Company+2011} and \cite{Dominguez-Sanchez+2018} are plotted, respectively, with solid red and dashed cherry lines.
The differences seen are actually smaller than those found when applying the ALFALFA selection, and hence they would be considered as second order effect.
We conclude that our inferences may depend on the assumed morphological classification --as well as on the separation criterion for defining LTG and ETGs-- but the main trends and conclusions are unaffected by this.

\subsubsection {Errors in membership and central/satellite designation}

As discussed in Section \ref{sec:HI_mass}, we assign \HI\ masses to halos and subhalos {\it identified in the simulation} by using the \RHI\ conditional distributions derived in Paper II and separated into centrals and satellites following the prescription from \cite{Calette+2021}. For such prescription, in that paper we used the \texttt{xGASS} \citep{Catinella+2018} survey and the central/satellite designation tabulated for each galaxy. As  shown in \citet[][see also \citealp{Bravo-Alfaro+2000}]{Campbell+2015}, group finders may suffer from membership allocation and central/satellite designation errors.
 Therefore, a potential concern is the impact of errors from group finders in the central/satellite \RHI\ conditional distributions reported in \citet{Calette+2021}. As is described in \citet{Campbell+2015}, errors from group finders tend to smooth out the differences between centrals and satellites, which leaves the possibility that the differences between their conditional distributions are larger to what was reported in \citet{Calette+2021}. We notice, however, that the \texttt{xGASS} team improved the \citet{Yang+2007} group membership by visually inspecting false pairs and galaxy shredding \citep[see][]{Janowiecki+2017}. Thus, we expect that the \citet{Calette+2021} \HI\ conditional distributions will be only marginally affected by errors in central/satellite designation.

\subsubsection{The comparison to the ALFALFA \HI\ clustering}
\label{sec:Caveats-clustering}

In Section \ref{secc:clustering}, we showed that
our mock galaxy catalogue tends to overpredict the 
observed \HI\ clustering from the ALFALFA survey, even after 
imposing an ALFALFA-like selection 
effect and in the extreme case of assuming that galaxies
in this survey consist only of late-types. Thus, a potential concern 
is in our method when assigning morphology
to the galaxies seeded in the simulation haloes. 
As explained in Section \ref{sec:Morphologies}, we assumed 
that the scatter around the \ms--\vmax\ relation is
independent of galaxy morphology. We are aware that 
this assumption implies that clustering properties of 
ETGs and LTGs are almost identical, at least for the two-halo
term. Thus, 
by construction one expects that our mock catalogue will not
recover the well known clustering properties 
of ETGs and LTGs as a function of stellar mass.
This shortcoming in combination with the ALFALFA selection effects
may work in the direction of overpredicting the \HI\ clustering. 
There are at least three options that could help solve the above issue:
\begin{enumerate}
\item The scatter around the \ms--\vmax\ relation depends
    on morphology. In this assumption, ETGs and LTGs occupy haloes of different masses. In particular, if there is a strong correlation between halo and morphology in which haloes of ETGs are more massive than those of LTGs at fixed  \ms, then the clustering 
    of ETGs will be stronger due to the halo bias \citep[see e.g.,][]{Rodriguez-Puebla+2015, Correa_Schaye2020}.
    
\item Assembly bias. Halo masses of ETGs and LTGs could be identical
but their properties are different. 
If galaxy morphology strongly correlates with halo's assembly time, such as formation time, then the clustering 
    of ETGs will be stronger due to the halo assembly bias \citep[see e.g.,][]{Hearin_Watson2013, Correa_Schaye2020}.  

\item A combination of both.
\end{enumerate}

Studying the above options  
is beyond the scope of 
this paper, in which we have focused on studying the galaxy
population as a {\it whole}. While we will study in more detail
the dependence of the galaxy-(sub)halo connection with galaxy morphology 
in a forthcoming paper, this is actually
less important as here any dependence 
will be wash-out (averaged) when studying the whole population. 

Finally, the $\omega_p^{\rm HI}(r)$ functions for {\it all} galaxies 
measured in future large and deep \HI\ surveys, 
which will not be significantly affected by selection effects,
are expected to be 
similar to those reported in Section \ref{secc:clustering}.

\begin{figure}
\includegraphics[trim = 9mm 45mm 98mm 8mm, clip, width=1.05\columnwidth]{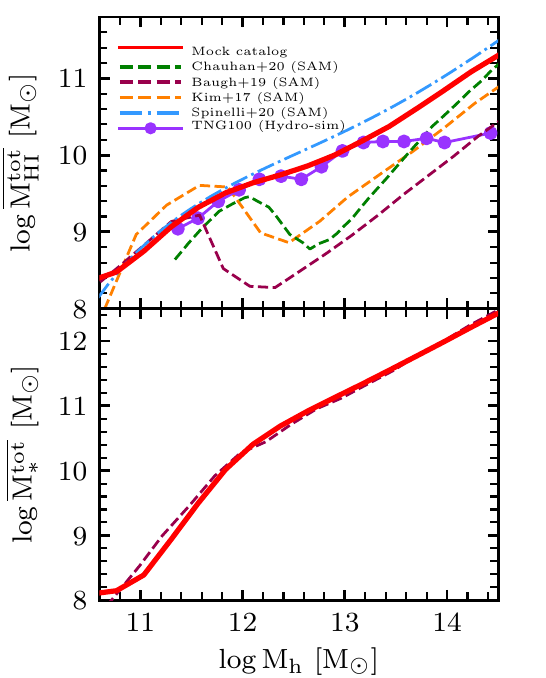}
\caption{ {\it Upper panel:} Median total \HI\ mass inside haloes of mass \mh\ as  empirically determined here (red line) compared to several predictions from SAMs and hydrodynamics simulations.
{\it Lower panel:} As upper panel but for the total stellar mass compared to the SAM results from
\citet{Baugh+2019}.
}
\label{fig:logMHI-logMs-logMh-tot-vs-theo-works} 
\end{figure}

\begin{figure*}
\includegraphics[trim = 9mm 75mm 45mm 5mm, clip, width=0.82\textwidth]{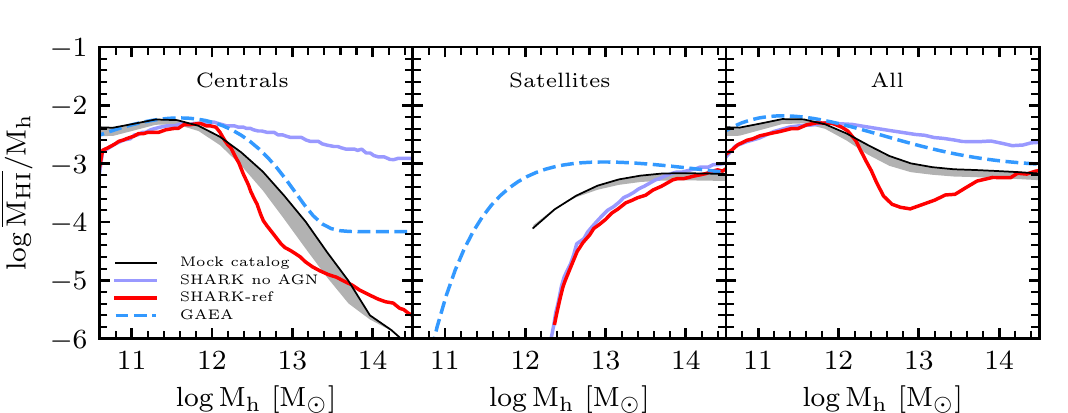}
\caption{ Median \HI-to-halo mass ratio as a function of \mh. From left to right we plot contributions of centrals, all satellites within the halo, and the sum of both. Results from our fiducial mock catalogue are presented as solid black lines and shaded areas; the line is for the \citet[][]{Huertas-Company+2011} morphological classification while the lower edge is for the \citet[][]{Dominguez-Sanchez+2018} classification. Red and purple lines correspond to the SHARK default and no AGN feedback models, respectively. Light blue dashed lines are results from GAEA.
} 
\label{fig:SHARK-comparison} 
\end{figure*}

\subsection{Comparisons to theoretical predictions}
\label{discussion:comparisons}

The new generation of SAM and cosmological hydrodynamics simulations, after post-processing, are able to predict the  total galaxy \HI\ gas content associated to haloes. In Figure \ref{fig:logMHI-logMs-logMh-tot-vs-theo-works} we show the \textit{median} of the total \HI\ mass,
$\overline{\mhatot}$, as a function of \mh\ for several SAMs \citep{Kim+2017,Baugh+2019,Spinelli+2020,Chauhan+2020} and the Illusttris-TNG100 simulation \citep{Stevens+2019,Chauhan+2020}. We have homogenized the halo masses to the virial mass. Our semi-empirical inference for the median $\overline{\mhatot}$--\mh\ relation is showed with the red thick line.

Our results show a relatively smooth increasing of $\overline{\mhatot}$ with \mh\ in the $3\times 10^{11}-10^{13}$ \msun\ mass range, similar to what is predicted by the GAEA SAM \citep{Spinelli+2020} and the Illustris-TNG100 simulation, but in tension with other SAMs: an old version of GALFORM \citep{Kim+2017}, its new version \citep{Baugh+2019}, and SHARK \citep{Chauhan+2020}. The decrease of $\overline{\mhatot}$ predicted by these models at the halo masses where the central dominates by far in the total \HI\ content of haloes ($\mh\lesssim 2\times 10^{12}$ \msun), is mainly associated with the AGN feedback. The feedback, mainly mechanical, keeps the gas in the halo hot, preventing it from accreting into the galaxy. 
The major difference is with GALFORM \citep{Baugh+2019}, even though we find excellent agreement with their total (central+satellites) stellar mass in haloes, see lower panel of Figure \ref{fig:logMHI-logMs-logMh-tot-vs-theo-works}. The AGN feedback strength affects more drastically the \HI\ gas content of galaxies than stellar masses in SAMs. Thus, {\it the \HI\ gas fraction of galaxies could be an important constrain for the way AGN feedback is included in models and simulations}, together with the more traditionally used high-mass end of the GSMF. 

Figure \ref{fig:SHARK-comparison} compares our results (black lines and shaded areas) with those of \citet[][SHARK]{Chauhan+2020} and \citet[][GAEA]{Spinelli+2020} regarding the median \HI\ gas content of centrals, of all satellites, and of the sum of both as a function of \mh, from left to right respectively. The red lines are for the default SHARK model while the purple ones are for the case of no AGN feedback. The SHARK \HI\ gas contents of centrals in haloes are only slightly lower than our semi-empirical determinations at the masses $\mh\lesssim 10^{12}$ \msun. For larger masses, ${\overline{M_{\rm HI}}}/\mh$ falls more sharply in the default SHARK model than our determinations, due to the effect of AGN feedback as discussed above. As for the sum of \HI\ gas in satellites within distinct haloes, SHARK predicts values significantly lower than our determinations, specially for $\mh<10^{13}$ \msun, and this contributes to the deep drop in the SHARK ${\overline{M^{\rm tot}_{\rm HI}}}/\mh$ ratio at halo masses in between $10^{12}$ and $5\times 10^{12}$ \msun. As shown in \citet[][]{Chauhan+2020}, the ram-pressure striping prescription (strangulation) has little effect on the \HI\ gas content of satellites. On the other hand, \citet[][]{Chauhan+2020} assume that the gas that is ejected from satellites due to stellar feedback is transferred to the ejected gas reservoir of the central. Therefore, that gas cannot be reincorporated into the hot halo gas of the satellites.  In the future, we will explore whether this assumption is or not relevant.

As for GAEA (dashed blue line), the \mha/\mh\ ratio of centrals agree well with our semi-empirical determinations. In particular, the drop at $\mh\gtrsim 10^{12}$ \msun, related to the AGN feedback strength, is only slightly less drastic than our determinations. However, at $\mh\gtrsim 3\times 10^{13}$ \msun, the \mha/\mh\ ratio becomes constant instead of showing a continuous drop as in our case or in the case of SHARK. Regarding the \HI\ contained in the satellites, GAEA predicts higher values than our determinations and much higher than those of SHARK. As a result, the total \HI-to-halo mass ratio in haloes is slightly higher in GAEA at $\mh\gtrsim 10^{12}$ \msun\ than our determinations, but is in line with the trend, showing no depression.

\section{Summary and Conclusions}
\label{conclusions}

In this paper we have presented a semi-empirical approach to link 
the \HI\ gas content of galaxies 
to their stellar masses and DM (sub)haloes in the \smdp\ N-body cosmological 
simulation. Galaxies are initially linked to 
DM (sub)haloes via the \ms--\vmax\ relation\footnote{ 
Here we assumed that \vmax\ corresponds to the maximum 
circular velocity for distinct haloes, whereas for subhaloes 
it is the peak maximum circular velocity 
reached along the halo’s main progenitor branch, see Eq. (\ref{vmax-def}).} 
derived from SHAM. We assume that 
galaxies are lognormally distributed around the 
\ms--\vmax\ relation with a dispersion of 0.15 dex. 
To every galaxy in the catalogue we assign either an 
early- or late-type morphology based on the
observed fractions of ETGs for centrals and satellites 
from the SDSS \citet{Yang+2012} group catalogue and the \citet{Huertas-Company+2011}
morphological classification of SDSS galaxies. Finally, we assign a \HI\ mass
using the central/satellite \mha{} conditional PDFs given \ms\
of ETGs and LTGs derived from observations in Paper I and in \citet{Calette+2021}. 
Thus, for every halo or subhalo in the 
\smdp\ simulation we have assigned a stellar mass, 
galaxy morphology (LTG or ETG), and \HI\ mass. We emphasize that the morphology
assignation is a {\it necessary } step for sampling
the empirical \mha{} distributions, though in this paper, we 
are interested in studying the galaxy stellar-\HI-halo 
connection for the galaxy population as a {\it whole} and not in
its segregation by morphology. 
Our main results are as follows:

\begin{itemize}

    \item The value of $\langle\log\mha\rangle$ for the whole population as a function of \vmax\ (\mhp) monotonically increases up to $\vmax\sim 160$ km/s ($\mhp\sim 10^{12}$ \msun) where it reaches a maximum value of $\langle\log(\mha/\msun)\rangle \sim 9.2$, Fig. \ref{fig:HI-halo-relations}. 
    At higher (sub)halo velocities (masses) it decreases only slightly. This is in contrast to  $\langle\log\ms\rangle$ which increases monotonically as a function of \vmax\ (\mhp).

    \item The scatter around $\langle\log\mha\rangle$ increases with \vmax\ (\mhp). At low (sub)halo velocities (masses), $\vmax\sim80$km/s ($\mhp\sim10^{11}\msun$) this is $\sim0.5$ dex. At $\vmax\gtrsim 160$ km/s ($\mhp\gtrsim 10^{12} \msun$) 
    it increases rapidly reaching a maximum
    of $\approx1.2$ dex. 
    
    \item In general, the \mha\ conditional PDFs as a function of \vmax\ (\mhp) are broad and highly asymmetric with a long tail towards lower values of \mha. For $\vmax\gtrsim 160$ km/s ($\mhp\gtrsim 10^{12} \msun$), the PDFs are bimodal with a second peak appearing at lower values of \mha, a behaviour
    simply inherited from the input \mha\ conditional PDFs as function of \ms.  
    As a result, different statistical estimators as logarithmic mean, arithmetic mean, and median of \mha{} as a function of \vm\ (\mhp) differ among them (Fig. \ref{fig:HI-halo-stats}). 
    
    \item There are not significant differences in
    $\langle\log\mha\rangle$ as a function of \vmax\ or \mhp\ between centrals and satellites (recall that the velocity or mass of subhaloes is calculated at their peak value in the past). On average, satellites have slightly lower values of \HI\ mass at a given \vmax\ or \mhp\ than centrals (Fig. \ref{fig:HI-halo-relations}). The  scatter is broader for satellites than for centrals by 0.25 dex for $\vmax\lesssim 160$ km/s and 0.3 dex for $\mhp\lesssim 10^{12}$ \msun.

    \item The \HI\ projected 2PCFs from our mock catalogue  increase only slowly in amplitude with  increasing \HI\ mass threshold (Fig. \ref{fig:hi_2pcf_mock}). 
    
    \item Assuming identical \mha\ mass conditional PDFs for centrals and satellites results in 2PCFs that have higher amplitudes, specially at the 1-halo term, than in our more realistic fiducial model, where satellites have lower \HI\ gas fractions than centrals, see  Figs. \ref{fig:RHI-Mh} and \ref{fig:hi_2pcf_mock}.
    
    \item Our predicted \HI\ projected 2PCFs have higher amplitudes than the ones measured in the blind \HI\ ALFALFA survey, see Fig. \ref{fig:HI-2PCFs}. When emulating ALFALFA selection effects in our catalogue, which selects gas-rich late-type galaxies, the projected 2PCFs are closer, but yet slightly above ALFALFA particularly for $\mha>10^{10}$ \msun. 
    
    \item The simple SHAM is unable to reproduce realistic 2PCFs  (Fig. \ref{fig:HI-2PCFs-vs-am}) as it assumes identical $\langle\log\mha(\vmax)\rangle$ relations for centrals and satellites and predicts that it monotonically increasing.   

    \item The total galactic (central + satellites) \HI\ gas content in distinct haloes, \mhatot, strongly depends on \mh\ and is completely dominated by centrals up to $\sim 10^{12}$ \msun, while at higher masses the total contribution from satellites increases; at $\mh\sim 10^{13}$ \msun\ both contributions are roughly equal and at $\mh=10^{14}$ \msun\ the total \HI\ in satellites is larger by $>1$ dex than the central galaxies contribution (Fig. \ref{fig:logMHI-logMh-tot-cen-Sumsat}). The inferred mean \mhatot--\mh\ relation does not present a dip at the masses where AGN feedback is expected to be important. 
\end{itemize}

The results presented here offer new constraints on the empirical galaxy-(sub)halo connection both for stellar and \HI\ masses.  These results are relevant for calibrating and testing the predictions from new generation SAM and hydrodynamics simulations of galaxy evolution within the $\Lambda$CDM cosmology. 
We have presented some preliminary comparisons with some of these theoretical predictions and discussed that our semi-empirical results can be relevant to constrain the strength of AGN feedback and the environmental processes that satellite galaxies undergo.

Our results on the galaxy \HI-(sub)halo connection show that 
\mha\ is not totally determined by the halo properties $\vm$ or $\mhp$.
The above has deep implications as it infringes the
central assumptions in the galaxy-halo connection models
such as the HOD and SHAM. Thus, alternative tools, like the ones used here, should be adopted in order to derive a realistic \HI-to-halo mass relation.
We also found that the \HI\ conditional PDFs as a function of \vm\ or \mhp\ are highly asymmetric and even bimodal. 
In view of the above, to characterize the dependence of \mha\ on \mhp\ at large masses is better to use the median of \mha\ as a function of \mhp\ (or \vmax).
For the median \mha, we see evidence of a weak  decrease of \mha\ with \mhp\ for (sub)haloes larger than $\mhp\sim10^{12}$ \msun, suggesting that the galaxies in these (sub)haloes have exhausted most of their gas reservoir, have not accreted in a long time or have even ejected their gas, e.g. due to AGN feedback.

On the other hand, we presented predictions on the \HI\ spatial clustering for the whole population of local galaxies. Currently,  blind \HI\ surveys, such as ALFALFA, are shallow and introduce strong selection effects to some properties of their optical hosts. We have shown that while these selection effects are not crucial for measuring the \HI\ mass function (see Fig. \ref{fig:HIMF-all-cen-sat}), they become critical for measuring the \HI\ spatial clustering (as well as for the \HI-to-stellar mass correlation, see Papers I and II). In this sense, we showed that using the ALFALFA \HI\ spatial clustering for constraining the \HI\ galaxy-halo connection may lead to incorrect results, unless the biases described above have been introduced within the galaxy-halo connection model employed. The results on the \HI\ spatial clustering presented here can be used for testing theoretical predictions as well as for estimates in the designing of larger and deeper \HI\ surveys that will be completed with forthcoming radio telescopes such as SKA or its Pathfinder instruments. 

Finally, the fact that our mock catalogue predicts (slightly) higher \HI\  clustering than ALFALFA even after imposing the selection effects of ALFALFA, suggests that the scatter around the \ms--\vmax\ relation should depend on morphology in such a way that the clustering properties of LTGs and ETGs differ, as observations show. In a forthcoming paper we will explore the effects of introducing the above-mentioned dependence of the scatter on morphology. In any case, we highlight that the results presented here for the \textit{whole} galaxy population are valid whether there is a dependence of the scatter around the \ms--\vmax\ relation on morphology or not.

\section*{Acknowledgements}

We thank the anonymous referee for the comments and suggestions that helped improve the presentation of this paper.
ARC acknowledges CONACyT for a PhD fellowship. ARP and VAR acknowledge financial support from CONACyT through ``Ciencia Basica'' grant 285721, and  from DGAPA-UNAM through PAPIIT grant IA104118.
CL has received funding from the ARC Centre of Excellence for All Sky Astrophysics in 3 Dimensions (ASTRO 3D), through project number CE170100013.

\section*{Data availability}
The data underlying this article will be shared on reasonable request to the corresponding author.



\bibliographystyle{mnras}
\bibliography{references} 



\appendix
\section{Tables for the galaxy-halo connection}\label{data-tables}

Table \ref{table-main-VDM} presents the data related to the bottom-left panel of Figure \ref{fig:HI-halo-relations}. For different values of $\log(\vmax/{\rm km s^{-1})}$, we give the corresponding  $\langle\log(\mha/\msun)\rangle$ values and their standard deviations for all, central, and satellite galaxies.  

Table \ref{table-main-logMHI-logMD} presents the data related to the bottom-right panel of Figure \ref{fig:HI-halo-relations} and to Figure \ref{fig:logMHI-logMh-tot-cen-Sumsat}. For different values of $\log(\mhp/\msun)$ showed in column (1), we give the corresponding $\langle\log(\mha/\msun)\rangle$ values and their standard deviations (in dex) for all, central, and satellite galaxies, columns from (2) to (7), respectively. Columns (8) and (9) are the mean arithmetic values of the total \HI\ mass and the sum of the \HI\ mass in satellites inside distinct haloes, respectively (upper panel of Fig. \ref{fig:logMHI-logMh-tot-cen-Sumsat}). Columns (10) and (11)  are as columns (8) and (9) but for stellar masses, respectively (lower panel of Fig. \ref{fig:logMHI-logMh-tot-cen-Sumsat}).

\begin{table*}
	\centering
	\Large
	\caption{Mean logarithmic $\mha$--$\vmax$ relations and their standard deviations for all, central, and satellite galaxies as presented in Fig. \ref{fig:HI-halo-relations}.}  
	\resizebox{12cm}{!} {
		\begin{tabular}{ccccccc}
			\hline
			\hline
			$\log{V_{\rm DM}}$  & $\langle\log{M_{\rm HI}}\rangle$   &  $\sigma$ & $\langle\log{M_{\rm HI}}\rangle_{c}$   &  $\sigma_{c}$ & $\langle\log{M_{\rm HI}}\rangle_{s}$   &  $\sigma_{s}$   \\   
			
			(1)  & (2) & (3) & (4) &  (5) & (6) &  (7) \\   \hline

    1.80 &     8.09  &   0.59  &   8.30 &    0.43  &   7.54 &   0.57  \\
1.89 &    8.40  &   0.64  &   8.66 &   0.46  &   7.92 &   0.64  \\
1.99 &    8.77  &   0.68  &   9.02 &   0.50  &    8.35 &   0.73  \\
   2.08 &     9.04  &   0.75  &   9.25 &   0.60  &   8.68 &   0.84  \\
2.17 &      9.15  &   0.88  &   9.32 &    0.77  &   8.85 &   0.99  \\
2.27 &    9.10  &    1.03  &   9.23 &   0.96  &   8.85 &    1.11  \\
   2.36 &    9.01  &    1.11  &   9.10 &    1.08  &   8.81 &    1.15  \\
2.45 &    8.91  &    1.15  &   8.96 &    1.14  &   8.76 &    1.15  \\
2.55 &    8.84  &    1.15  &   8.87 &    1.15  &   8.74 &    1.12  \\
   2.64 &    8.76  &    1.13  &   8.79 &    1.15  &   8.68 &    1.07  \\
2.73 &    8.75  &    1.10  &   8.77 &    1.12  &   8.64 &    1.01  \\
2.83 &    8.71  &    1.08  &   8.73 &    1.09  &   8.58 &   0.99  \\
   2.92 &     8.72  &    1.07  &   8.74 &    1.09  &   8.60 &   0.86  \\
3.01 &    8.77  &    1.04  &   8.77 &    1.06  &   8.77 &    0.78  \\
3.11 &    8.77  &   0.94  &    8.77 &   0.95  &   8.78 &   0.66  \\
    3.20 &    9.12  &   0.92  &   9.12 &   0.94  &   9.07 &    0.11        
   \\ \hline		
\end{tabular}
	}
	\label{table-main-VDM}
\end{table*}

\begin{table*}
	\centering
	\Large
	\caption{Mean logarithmic $\mha$--$\mhp$ relations and their standard deviations for all, central, and satellite galaxies (columns 1--7), as plotted in Fig. \ref{fig:HI-halo-relations}, and mean arithmetic \mhatot--\mh, $M^{\rm sat}_{\rm HI}$--\mh, $M^{\rm tot}_{\rm \ast}$--\mh, and $M^{\rm sat}_{\rm \ast}$--\mh\ relations (columns 8--11, respectively), as plotted in Fig. \ref{fig:logMHI-logMh-tot-cen-Sumsat}.
	}  
	\resizebox{18cm}{!} {
		\begin{tabular}{ccccccccccc}
			\hline
			\hline
			$\log\mhp$  & $\langle\log\mha\rangle$   &  $\sigma$ & $\langle\log\mha\rangle_{c}$   &  $\sigma_{c}$ & $\langle\log\mha\rangle_{s}$   &  $\sigma_{s}$ & $\log\langle{\rm M}^{\rm tot}_{\rm HI} \rangle$ & $\log\langle{\rm M}^{\rm sat}_{\rm HI}\rangle$ & $\log\langle{\rm M}^{\rm tot}_{\rm \ast} \rangle$ & $\log\langle{\rm M}^{\rm sat}_{\rm \ast}\rangle$  \\   
			(1)  & (2)   &  (3) & (4)   &  (5) & (6)   &  (7) & (8) & (9) & (10) & (11)  \\   \hline
			
10.5 &    7.97  &    0.56  &   8.15 &   0.43  &   7.58    &   0.61  &   8.49   &  4.88  &   8.16    &  5.80   \\
10.8 &    8.26  &   0.56  &   8.38 &   0.46  &   7.95     &   0.69  &   8.67   &  5.62  &   8.27    &  6.10   \\
11.1 &     8.64  &   0.60  &   8.75 &    0.49  &   8.36   &   0.76  &   9.00   &  6.51  &   8.66    &  6.74   \\
11.4 &     8.97  &   0.66  &   9.07 &   0.53  &   8.67    &   0.86  &   9.32   &  7.28  &   9.27    &  7.61   \\
11.7 &    9.16  &   0.76  &   9.27 &   0.64  &   8.83     &   0.99  &   9.57   &  7.91  &   9.85    &  8.32   \\
12.0   &  9.20    & 0.91    & 9.30   & 0.82    & 8.85       &   1.09  &   9.74   &  8.46  &   10.32   &  8.98   \\
12.3 &    9.12  &     1.04  &   9.20 &   0.99  &   8.82   &   1.14  &   9.87   &  8.94  &   10.66   &  9.57   \\
12.6 &    9.01  &    1.11  &   9.07 &    1.10  &   8.77   &   1.15  &   9.99   &  9.34  &   10.92   &  10.08  \\
12.9 &     8.91  &    1.14  &    8.94 &    1.14  &   8.74 &   1.13  &   10.13  &  9.70  &   11.15   &   10.52 \\
13.2 &    8.82  &    1.15  &   8.85 &    1.16  &   8.68   &   1.08  &   10.30  &  10.04 &   11.38   &  10.92  \\
13.5 &    8.77  &    1.12  &   8.78 &    1.14  &   8.67   &   1.01  &   10.49  &  10.34 &   11.62   &  11.28  \\
13.8 &    8.72  &    1.11  &   8.75 &    1.12  &   8.52   &   0.98  &   10.74  &  10.65 &   11.87   &  11.61  \\
14.1 &    8.73  &    1.07  &   8.74 &    1.08  &   8.61   &   0.94  &   10.98  &  10.93 &   12.11   &  11.92  \\
14.4 &    8.81  &    1.09  &   8.80 &    1.11  &   8.83   &   0.88  &   11.24  &  11.21 &    12.36  &   12.21 \\
14.7 &    8.70  &    1.01  &   8.70 &    1.02  &   8.70   &   0.69  &   11.50  &  11.47 &   12.61   &  12.49  \\
15.0   &  8.95    & 0.89    & 8.96   & 0.91    & 8.77   & 0.35  &   11.70  &  11.69 &   12.80   &   12.71     
   \\ \hline		
\end{tabular}
	}
	\label{table-main-logMHI-logMD}
\end{table*}

\label{lastpage}
\end{document}